%% file: iclr2025_conference.tex
\documentclass{article} % For LaTeX2e
\usepackage{iclr2025_conference,times}

\input{math_commands.tex}

\usepackage{hyperref}
\usepackage{url}
\usepackage{booktabs}       % professional-quality tables
\usepackage{amsfonts}       % blackboard math symbols
\usepackage{nicefrac}       % compact symbols for 1/2, etc.
\usepackage{microtype}      % microtypography
\usepackage{xcolor}         % colors
\usepackage{wasysym}
\usepackage{graphicx}
\usepackage{tabulary}
\usepackage{pifont}
\usepackage{subfigure}
\usepackage{textcomp}
\usepackage{multirow}

\title{RecFlow: An Industrial Full Flow Recommendation Dataset}

\author{Qi Liu$^1$, \And Kai Zheng$^2$, \And Rui Huang$^2$, \And Wuchao Li$^1$, \And Kuo Cai$^2$, \And Yuan Chai$^2$, \And Yanan Niu$^2$, \And Yiqun Hui$^2$,\And Bing Han$^2$,\And Na Mou$^2$, \And Hongning Wang$^4$, \And Wentian Bao$^3$, \And Yunen Yu$^3$, \And Guorui Zhou$^2$, \And Han Li$^2$, \And Yang Song$^2$, \And Defu Lian$^1$, \And Kun Gai$^3$ \\
$^1$University of Science and Technology of China \quad  $^2$Kuaishou\\
$^3$Independent \quad $^4$Tsinghua University\\
\texttt{\{qiliu67,liwuchao\}@mail.ustc.edu.cn,} \quad \texttt{\{liandefu\}@ustc.edu.cn}\\
\texttt{\{zhengkai,huangrui06,caikuo,niuyanan,chaiyuan\}@Kuaishou.com}\\
\texttt{\{huiyiqun,hanbing,zhouguorui,lihan08,songyang\}@Kuaishou.com} \\
\texttt{\{hw5x\}@virginia.edu,} \quad \texttt{\{wb2328\}@columbia.edu} \\
\texttt{\{yuenyun\}@126.com,} \quad \texttt{\{285208254,gai.kun\}@qq.com}
}

\iclrfinalcopy % Uncomment for camera-ready version, but NOT for submission.
\begin{document}

\maketitle

\begin{abstract}
Industrial recommendation systems (RS) rely on the multi-stage pipeline to balance effectiveness and efficiency when delivering items from a vast corpus to users. Existing RS benchmark datasets primarily focus on the exposure space, where novel RS algorithms are trained and evaluated. However, when these algorithms transition to real-world industrial RS, they face a critical challenge: handling unexposed items—a significantly larger space than the exposed one. This discrepancy profoundly impacts their practical performance. Additionally, these algorithms often overlook the intricate interplay between multiple RS stages, resulting in suboptimal overall system performance. To address this issue, we introduce RecFlow—an industrial full-flow recommendation dataset designed to bridge the gap between offline RS benchmarks and the real online environment. Unlike existing datasets, RecFlow includes samples not only from the exposure space but also unexposed items filtered at each stage of the RS funnel. Our dataset comprises 38M interactions from 42K users across nearly 9M items with additional 1.9B stage samples collected from 9.3M online requests over 37 days and spanning 6 stages. Leveraging the RecFlow dataset, we conduct courageous exploration experiments, showcasing its potential in designing new algorithms to enhance effectiveness by incorporating stage-specific samples. Some of these algorithms have already been deployed online, consistently yielding significant gains. We propose RecFlow as the first comprehensive benchmark dataset for the RS community, supporting research on designing algorithms at any stage, study of selection bias, debiased algorithms, multi-stage consistency and optimality, multi-task recommendation, and user behavior modeling. The RecFlow dataset, along with the corresponding source code, is publicly available at \textcolor{red}{\url{https://github.com/RecFlow-ICLR/RecFlow}}. The dataset is licensed under CC-BY-NC-SA-4.0 International License.
\end{abstract}

\section{Introduction}

Recommendation systems (RS) play a pivotal role in modern web and mobile applications that handle vast amounts of information. Their primary objective is to deliver personalized recommendations from an extensive corpus of items, based on estimated user preferences. To meet stringent online latency requirements, industrial RS predominantly employs a multi-stage funnel-like pipeline~\citep{covington2016deep}, striking a balance between effectiveness and efficiency. Substantial efforts have been devoted to designing algorithms within this system, aiming to enhance its effectiveness as measured by user feedback on selected items. A typical multi-stage RS consists of successive stages: \textbf{retrieval $\rightarrow$ pre-ranking $\rightarrow$ ranking $\rightarrow$ re-ranking}. 
During online serving, the retrieval stage~\citep{hidasi2015session,kang2018self,zhu2018learning} retrieves thousands of preferred items from the entire corpus. 
% Retrieval models~\citep{hidasi2015session,kang2018self,zhu2018learning} are primarily trained using items with positive feedback as positive samples and randomly sampled items from the corpus as negative samples. 
The pre-ranking stage~\citep{huang2013learning,wang2020cold} filters out less favorable items from the retrieved set, forwarding hundreds of more promising items to the ranking stage. In turn, the ranking stage~\citep{cheng2016wide,zhou2018deep,bian2022can} selects the most appealing items from this refined set. Finally, the re-ranking~\citep{pei2019personalized,bello2018seq2slate} stage determines the final items to be displayed, considering both diversity and business objectives. Notably, as we progress through the stages, the model complexity tends to increase, incorporating additional features and interleaving them at shallow layers of deep neural network models. 
Importantly, the latter three stages typically learn from the exposure space, which captures actual user feedback (both positive and negative) on the displayed items.

While the mature industrial RS paradigm performs adequately, it still faces two significant shortcomings. First, a discrepancy exists between the data distribution in the training space and that in the serving space~\citep{qin2022rankflow}. The former corresponds to the exposed space, while the latter primarily resides in the unexposed space. This discrepancy, which we refer to as the distribution shift problem, poses challenges. For instance, consider the pre-ranking model~\citep{wang2020cold}: it must score thousands of items, yet only a few of these items are exposed to users and stored as training data in each request. Most of the remaining samples have not been exposed even once. Consequently, a pre-ranking model trained solely on the exposure space may inaccurately predict preferences in the retrieved space, leading to suboptimal recommendations~\citep{wei2024enhancing}. Similar issues arise in the ranking and re-ranking stages. Second, there is a discrepancy between the learning and serving environments. Although models at different stages are learned and evaluated separately, they must collaborate as a cohesive system to meet user preferences. Insufficient knowledge about other stages during the learning process can result in suboptimal performance when these learned models serve online. For example, the online performance of a retrieval algorithm not only depends on its own characteristics but is also influenced by subsequent stages. Incorporating knowledge from these subsequent stages can potentially enhance the retrieval algorithm’s performance~\citep{ding2019reinforced,lou2022re,zheng2024full}. 

As we are aware, large-scale datasets serve as the bedrock for advancing various machine learning algorithms. For instance, ImageNet~\citep{deng2009imagenet} has significantly contributed to computer vision, while GLUE~\citep{wang2018glue} has played a crucial role in natural language processing. However, in the RS domain, existing RS datasets~\citep{harper2015movielens,ni2019justifying,asghar2016yelp,zhu2018learning,yuan2022tenrec,gao2022kuairec,gao2022kuairand,sun2023kuaisar}—though instrumental in fueling RS research—have a limitation: they are exclusively collected from the exposure space. Consequently, these datasets cannot fully capture the true dynamics of online recommendation services. Moreover, this inherent bias prevents them from effectively addressing the discrepancy between training and serving in RS.

To address this issue, we propose RecFlow, an industrial large-scale full-flow dataset collected from the real industrial RS. The industrial RS's multi-stage funnel-like pipeline encompasses the following stages: retrieval, pre-ranking, coarse ranking, ranking, re-ranking, and edge ranking. Unlike all previous RS benchmarks, RecFlow samples representative unexposed items from each stage of the funnel in a single request, alongside all the exposed items. The inclusion of full-stage samples in each request provides several merits: (1) By recording items from the serving space, RecFlow enables the study of how to alleviate the discrepancy between training and serving for specific stages during both the learning and evaluation processes~\citep{qin2022rankflow}. (2) RecFlow also records the stage information for different stage samples, facilitating research on joint modeling of multiple stages, such as stage consistency or optimal multi-stage RS~\citep{zheng2024full}. (3) The positive and negative samples from the exposure space are suitable for classical click-through rate prediction or sequential recommendation tasks~\citep{zhou2018deep,kang2018self}. (4) RecFlow stores multiple types of positive feedback (e.g., effective view, long view, like, follow, share, comment), supporting research on multi-task recommendation~\citep{ma2018modeling,zhao2019recommending,tang2020progressive,liu2023deep}. (5) Information about video duration and playing time for each exposure video allows the study of learning through implicit feedback, such as predicting playing time~\citep{covington2016deep,lin2023tree}. (6) RecFlow includes a request identifier feature, which can contribute to studying the re-ranking problem~\citep{pei2019personalized,bello2018seq2slate}. (7) Timestamps for each sample enable the aggregation of user feedback in chronological order, facilitating the study of user behavior sequence modeling algorithms~\citep{zhou2018deep,zhou2019deep,chang2023twin,hou2023deep}. (8) RecFlow incorporates context, user, and video features beyond identity features (e.g., user ID and video ID), making it suitable for context-based recommendation~\citep{huang2019fibinet,wang2022enhancing}. (9) The rich information recorded about RS and user feedback allows the construction of more accurate RS simulators or user models in feed scenarios~\citep{shi2019virtual,zhao2023kuaisim}. (10) Rich stage data may help estimate selection bias more accurately and design better unbiased algorithms~\citep{chen2023bias}. Furthermore, RecFlow is a large-scale dataset, containing 38 million exposure samples and 1.9 billion stage samples, ensuring the credibility of algorithm improvements based on its data.

Given these characteristics, RecFlow can be utilized across a broad spectrum of RS algorithms. In this paper, we primarily conduct pioneering experiments to explore its potential in each stage of the RS funnel. In the retrieval stage, we investigate the effectiveness of using filtered videos from each stage as hard negative samples and explore the interplay between retrieval and subsequent stages. For the coarse ranking stage, we leverage corresponding stage samples to address the distribution shift problem and model mutual effects between stages. Motivated by existing works, we explore how to exploit stage samples for designing auxiliary ranking tasks and behavior sequence modeling algorithms to improve classical AUC metrics. Similar exploration experiments are also conducted for the ranking stage. Notably, RecFlow also introduces a new recall metric to assess the performance of different methods based on stage samples to mitigate the gap between training and serving environment. To the best of our knowledge, RecFlow is the first RS dataset containing stage samples. It stands as one of the largest and most comprehensive datasets for RS, covering nearly all recommendation tasks. We have made the dataset and source codes publicly available to promote reproducibility and advance RS research.

\section{Dataset Characteristic}

\begin{figure*}
    \centering    
    \includegraphics[width=1.0\textwidth]{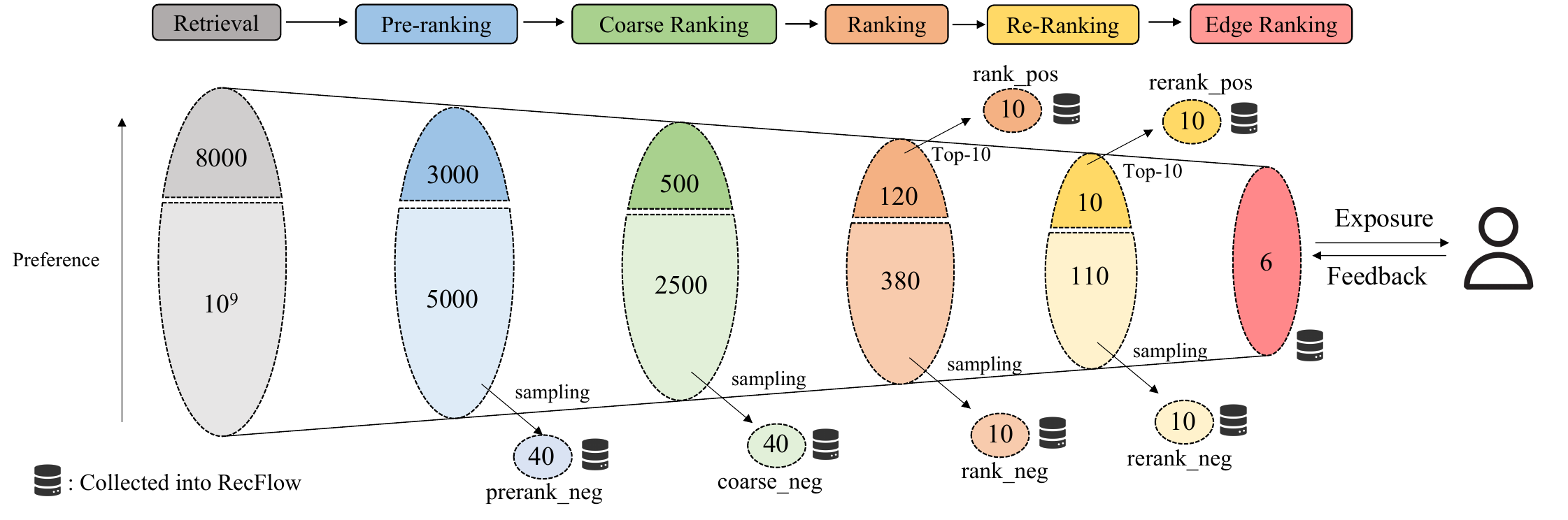}
    \caption{The overall collection process of RecFLow.}
    \label{fig:recflow}
\end{figure*}

\subsection{Collection}
RecFlow is the first RS dataset containing intermediate filtered videos of each stage in the industrial RS funnel. The multi-stage funnel-like pipeline of the industrial RS contains six stages including \textbf{retrieval $\rightarrow$ pre-ranking $\rightarrow$ coarse ranking $\rightarrow$ ranking $\rightarrow$ re-ranking $\rightarrow$ edge ranking}. The number of videos output at each stage is $8000\rightarrow3000\rightarrow500\rightarrow120\rightarrow10\rightarrow6$. We collect the online request logs from January 13 to February 18, 2024. The collection process is as follows. We randomly sample 42K seed users on January 12, 2024, and store each recommendation request of the seed users since January 13, 2024. As shown in Figure~\ref{fig:recflow}, we sample some filtered videos from each stage but adopt a stage-wise strategy. From January 13 to February 04, 2024, which is called the 1st period, we sample 10 filtered videos of the pre-ranking stage named pre-rank\_neg, 10 filtered videos of the coarse ranking stage named coarse\_neg, top 10 ranking videos as rank\_pos,  10 sampling filtered videos after the 120-th re-ranking video as rank\_neg in the ranking stage, top 10 re-ranking videos as rerank\_pos and 10 sampling filtered videos after the 80-th re-ranking video as rerank\_neg in the re-ranking stage, and the user's various feedbacks on the exposed videos. Note that the recommendation scenario is feeds-style, the user can only watch one video on the screen. So the 6 output videos of the RS may not all be exposed to the user because the user can leave the APP at any time. We define the realshow field to identify whether the user has watched the video. From February 05 to February 18, 2024, which is called the 2nd period, we expand the amount of stage samples. Both the pre-ranking\_neg and the coarse\_neg go up to 40. For the ranking, re-ranking, and edge ranking stages, we save all the videos that appear in these stages. We still obtain the rank\_pos, rank\_neg, rerank\_pos, rerank\_neg, and realshow under the same stage-wise strategy as the previous period. We collect stage samples in this way considering the storage pressure and information integrity. The 2nd period has more complete stage information compared to the 1st period, which gives the researchers more choices to further process the dataset based on their needs. We sample 10/40 filtered videos from the pre-ranking and coarse ranking stages because keeping all of the filtered videos has huge storage pressure. Besides, the videos filtered by the first three stages are less important. For the latter three stages, we keep the information integrity of the stage possibly. The videos appearing in these stages are closer to the user's preference and have a small scale.

\subsection{Features}
The formation of each instance in RecFlow is \{\textit{request\_id, request\_timestamp, user\_id, device\_id, age, gender, province, video\_id, author\_id, category\_level\_one, category\_level\_two, upload\_type, upload\_timestamp, duration, realshow, rerank\_pos, rerank\_neg, rank\_pos, rank\_neg, coarse\_neg, pre-rank\_neg, rank\_index, rerank\_index, playing\_time, effective\_view, long\_view, like, follow, forward, comment}\}. \textit{realshow} indicates whether the user has watched the video. The same procedure is applied to the other \textit{*\_pos/neg} fields. For example, when the video ranks top 10 in the ranking stage, then the \textit{rank\_pos} is set to 1 otherwise 0. To reserve the original industrial RS information, we also retain the ranking position of each video in the ranking and reranking stages through the \textit{rank\_index} and \textit{rerank\_index} fields respectively. We record seven types of positive feedback that reflect the user's varying degrees of preference towards videos. \textit{playing\_time} is the time the user spends watching the video. The other features' details are in the subsection Feature Description~\ref{feature_description} of Appendix. 

\subsection{Analysis}

\begin{table}[]
    \setlength{\tabcolsep}{3.6pt}
    \centering
    \small
    \caption{Detail quantity information of various aspects in RecFlow.}
    \begin{tabulary}{\textwidth}{CCCCCC}
    \toprule
            & \#Stage Sample & \#Request & \#Users & \#Realshow\_videos & \#All\_videos  \\
    \midrule
1st Period  & 352,120,401    & 6,062,348 & 38,193  & 5,984,924          & 30,305,725     \\
2nd Period & 1,572,217,303   & 3,308,233 & 35,073  & 3,627,694          & 55,665,503     \\
Total       & 1,924,337,704  & 9,370,581 & 42,472  & 8,773,147          & 82,216,301     \\
\toprule
            & \#Realshow & \#Like    & \#Long\_view & \#Effective\_view & \#Follow \\
\midrule
1st Period  & 24,523,473 & 1,027,013 & 5,853,054    & 9,343,776         & 69,495   \\
2nd Period & 13,721,842 & 618,158   & 3,111,439    & 5,063,751         & 37,558   \\
Total       & 38,245,315 & 1,645,171 & 8,964,493    & 14,407,527        & 107,053  \\
\toprule
            & \#Forward & \#Comment & \#Prerank\_neg & \#coarse\_neg & \#Rank\_pos  \\
\midrule
1st Period & 45,966     & 175,896    & 60,623,480     & 60,623,480     & 60,624,430   \\
2nd Period & 23,769     & 114,741    & 132,329,320    & 132,329,320    & 33,082,330   \\
Total       & 69,735    & 290,637   & 192,952,800    & 192,952,800    & 93,706,760   \\
\toprule
            & \#Rank\_neg & \#Rank        & \#Rerank\_pos & \#Rerank\_neg & \#Re-rank        \\
\midrule
1st Period  & 60,624,012  & 121,248,442   & 60,624,613   & 60,623,606   & 121,248,219    \\
2nd Period & 33,082,330  & 1,307,558,663 & 33,082,330   & 33,082,330   & 1,307,558,663  \\
Total       & 93,706,342  & 1,428,807,105 & 93,706,943   & 93,705,936   & 1,428,806,882  \\
\bottomrule
    \end{tabulary}
    \label{tab:statics}
\end{table}

\begin{figure}[htb]
	\centering
	\subfigure{
		\begin{minipage}{0.33\textwidth}
			\centering
			\includegraphics[width=\textwidth]{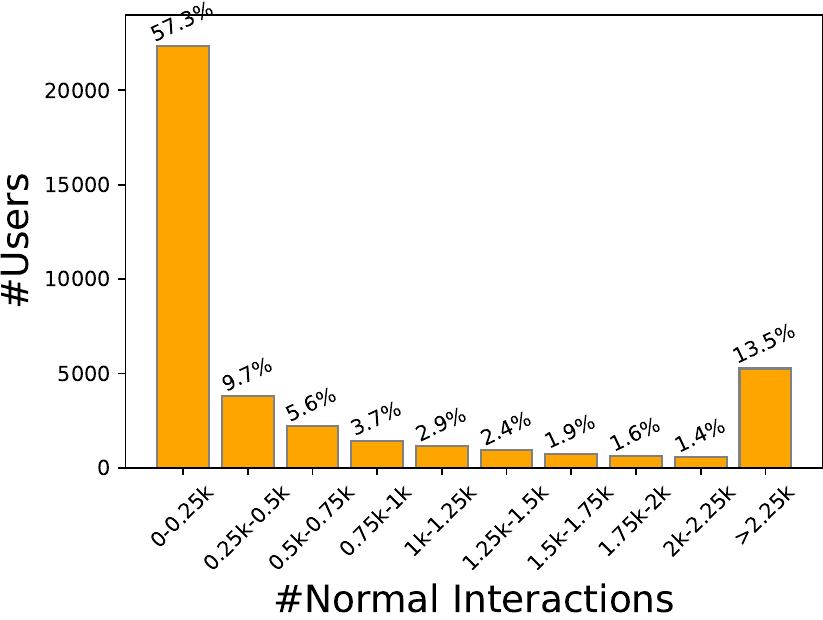}
			\caption{User Distribution.}
			\label{fig:User}
		\end{minipage}%
	}
		\hfill
	\subfigure{
		\begin{minipage}{0.635\textwidth}
			\centering
			\includegraphics[width=0.46\textwidth]{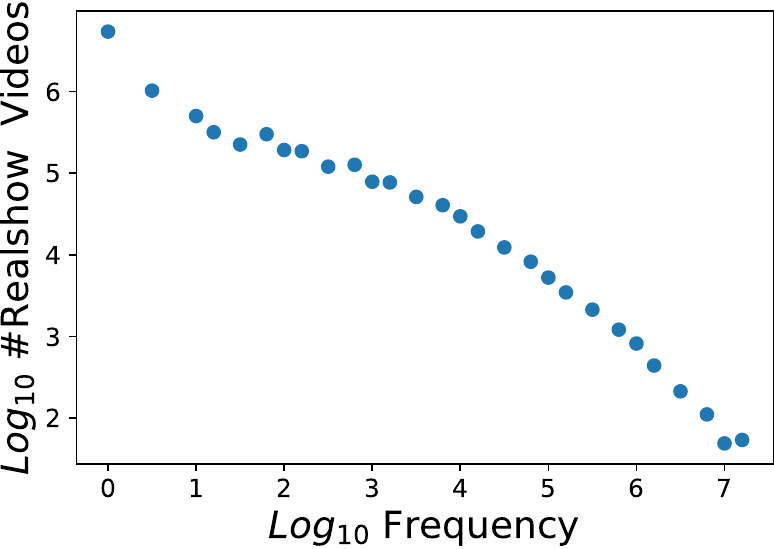}
			\includegraphics[width=0.46\textwidth]{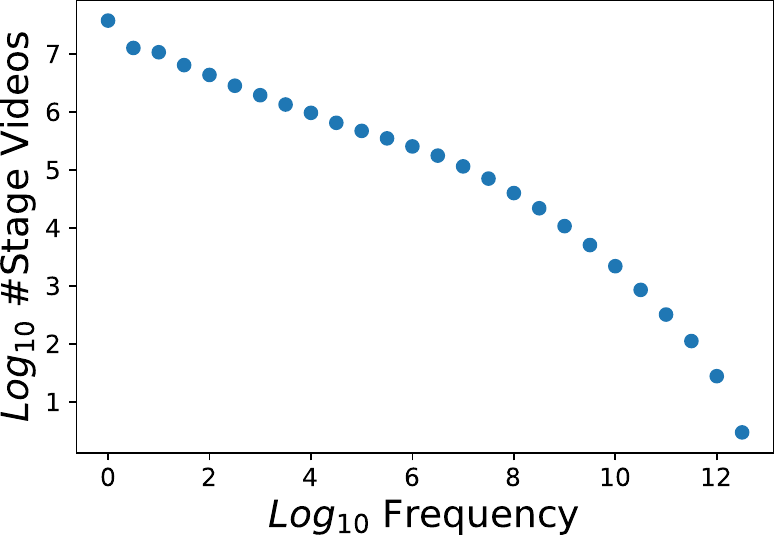}
                \vspace{1.2em}
			\caption{Video Distribution.}
			\label{fig:Item}
		\end{minipage}%
	}
\end{figure}

In this section, we conduct a basic statistical analysis to show RecFlow's characteristics. We collect 9 million requests. It has 38 million exposure samples and 1.9 billion stage samples (including exposure samples). Among these samples, there are 42K users, 8.7 million exposed videos, and 82 million videos. Nearly $89\%$ of videos are not exposed. This new character does not exist in existing RS datasets. During the first period, the quantity of each defined stage's samples is about 60 million. Stage samples are 14.8x larger than exposed samples. The difference between stage samples and exposure samples has increased to 236 times in the 2nd period. The huge quantity difference is the foundation for studying the distribution shift problem. The detailed quantities of the dataset are shown in Table~\ref{tab:statics}. Figure~\ref{fig:User}, whose horizontal axis represents the range of the number of videos interacted by users and the vertical axis shows the number and percentage of users within that range, illustrates that the frequency of users exhibits a long-tail distribution. In Figure~\ref{fig:Item}, the horizontal axis represents the logarithm of the frequency of video appearances, while the vertical axis shows the video quantity corresponding to that frequency. The left chart only includes videos marked as \textit{realshow} with 1, which are the exposed videos, while the right chart includes videos from all stages. It shows the frequency of videos in exposure space and all stages' space respectively. The left chart shows that exposure video frequency follows long-tail distribution. The right chart reveals that video frequency in all stages also obeys the long-tail distribution, which is not observed in previous datasets.  

\subsection{Comparison}
We state the characteristics of existing recommendation datasets to demonstrate the uniqueness of RecFlow. MovieLens~\citep{harper2015movielens} contains the user's rating data for movies. Amazon~\citep{ni2019justifying} dataset contains the user's review information on the product. Yelp~\citep{asghar2016yelp} is a dataset for location recommendation. The three datasets only contain the user's single type of positive feedback. Taobao~\citep{zhu2018learning}, an e-commerce dataset, has four types of the user's positive feedback. Tenrec~\citep{yuan2022tenrec} is a comprehensive recommendation dataset that has the characteristic of multiple types of the user's feedback from four different recommendation scenarios. KuaiRec~\citep{gao2022kuairec} is a full-observed video recommedation datatse. KuaiRand~\citep{gao2022kuairand} is an unbiased sequential video recommendation dataset with randomly exposed videos. KuaiSAR~\citep{sun2023kuaisar} is a unified search and recommendation dataset. The three datasets are opened for dedicated research problems. RecFlow differs from those datasets because of the existence of samples from each recommendation stage. Table~\ref{tab:dataset_comparison} in the subsection Dataset Comparison~\ref{dataset_comparison} of Appendix gives a detailed comparison between RecFlow and existing recommendation datasets.

\subsection{User consent and privacy protection}
We only collect interaction data from the user who has made his/her personal information publicly (like user\_id, age, gender, province, etc), and this public information allows for some level of data sharing, according to the privacy policy that users voluntarily agreed to when they signed up for an account. Besides, we anonymize all features that contain personal information. In detail, we anonymize each feature ID by adding the raw ID value with a random large integer first and remapping it to a new ID through the Hash algorithm. It can not know who is the person in the real world from the anonymous data. The General Data Protection Regulation of the European Union has confirmed that "personal information that has been anonymized does not belong to personal information. Therefore, personal information that has been anonymized does not have the corresponding personal information compliance obligations, and companies can freely process it without the consent of individuals." Thus, our open-source dataset meets legal requirements. 

We have anonymized all features which contain personal information including request\_id, user\_id, device\_id, age, gender, province, video\_id, author\_id, category\_level\_one, category\_level\_two, and upload\_type. We first anonymize each feature ID by adding the raw ID value with a random large integer and then remapping it to a new ID through the Hash algorithm. Note that each raw ID value owns a unique larger integer. The rest features are stage labels and the user's feedback labels, which are not related to privacy. Anonymizing data with random noise and the Hash algorithm satisfies the privacy protection requirements of the law of the European Union. The way of RecFlow's anonymization is more strict than previous public recommendation datasets including Amazon~\citep{ni2019justifying}, Taobao~\citep{zhu2018learning}, KuaiRec~\citep{gao2022kuairec}, and Tenrec~\citep{yuan2022tenrec}. We add random large integer noise before the Hash algorithm and others not. It is nearly impossible to recover raw personal information such as who is the person in the real world from anonymous data. 

\section{Experiments}
We explore how to utilize stage samples to alleviate distribution shift and distill knowledge of subsequent stages for improving RS's performance. We focus on the typical retrieval, coarse ranking, and ranking stages. For each stage, we briefly introduce its duty and existing learning paradigm. Then we state the motivation and the ways of exploiting stage samples. Finally, we report the experiment results and analysis. We run all experiments five times with Pytorch~\citep{imambi2021pytorch} on Nvidia 32G V100. We report the average result and standard deviation. For all methods and all experiments, we train the neural models for only one epoch and there is no early stopping. Thus, all methods are compared fairly. There are two reasons for only one epoch. First, all online recommendation models of the industrial RS are trained by one epoch. We keep consistency with the online configuration. Second, there exists one-epoch phenomenon~\citep{zhang2022towards} of training recommendation model which indicates that multi-epoch training does not bring improvement.

\subsection{Retrieval}
Retrieval is the first stage of the industrial RS. It aims at retrieving thousands of videos that the user potentially prefers from the 100 million scale video corpus. Given the large candidate pool, the retrieval stage mostly adopts the lightweight two-tower model together with approximate nearest neighbor search to retrieve items quickly. To ensure that the user's preferred videos are obtained, the retrieval models usually learn with positive feedback videos as positive samples and randomly sampling videos as negative samples. We choose SASRec~\citep{kang2018self} with one head and one layer for exploration experiments. We apply the effective\_view videos as positive samples and randomly sample $200$ videos as negative samples for each positive. To keep consistency with the real industrial RS's online learning mode, we train SASRec with the first $36$ days' data day by day. The data of the last day is for evaluation. We utilize the standard top-N ranking metrics including hit Recall@K and NDCG@K. K is set to $100$, $500$, $1000$. The feature is the user's $50$ past effective\_view videos. We apply embedding for the \textit{video\_id} feature and set the embedding dimension to $8$. The batch size is $4,096$ and the learning rate is $1e-1$. BPR~\citep{rendle2012bpr} is the loss function and Adam~\citep{kingma2014adam} is used for optimization. 

\subsubsection{Hard Negative Mining}
Recent research~\citep {zhang2013optimizing,rendle2014improving,lian2020personalized} have shown hard negative mining usually not only accelerates the convergence but also improves the model accuracy for the retrieval model. The hard negative samples are those videos that are similar to the positive videos but uninteresting to the user. The multi-stage RS pipeline aims at estimating the user's preference. Videos that fail to be exposed to the user during the pipeline are similar to the displayed positive video but very likely less attractive to the user. Thus, we think the unexposed stage samples indeed satisfy the definition of hard negative samples. We conduct experiments to explore the effectiveness of the stage samples as hard negative samples. In the experiments, we replace some randomly sampling easy negative samples with the same number of hard negative stage samples. The total number of negative videos for each positive video is $200$. 

We have the following findings from the result in Table \ref{tab:retrieval_hard}. (1) Applying filtered videos from each stage as hard negatives all gains performance improvement on the Recall/NDCG metric. (2) As the K in Recall/NDCG@K becomes smaller, the performance improvement becomes better. For example, when we add $1$ \textit{pre-rank\_neg} as hard negative, the relative promotion of Recall@100, 500, 100 are $24.7\%$, $18.2\%$, $9.2\%$ respectively, and the relative promotion of NDCG@100, 500, 100 are $28.3\%$, $20.7\%$, $12.6\%$ respectively. (3) The hard negative video from \textit{rerank\_pos} outperforms than the other stages. We think that videos from \textit{rerank\_pos} are negative samples of appropriate difficulty. We also vary the number of hard negative samples to observe the changes in the effectiveness. The experiment result and analysis are in the subsection~\ref{retrieval_num_hard_neg} of Appendix.

\begin{table}
\setlength{\tabcolsep}{2.5pt}
\renewcommand{\arraystretch}{1.0} 
\centering
\small
\caption{Recall(R) and NDCG(N) results (mean $\pm$ std) obtained by using a single different stage sample as the hard negative sample during the retrieval stage, with units of \%. The best and baseline results are based on the paired $t$-test at the significance level $5\%$.}
\label{tab:retrieval_hard}
\begin{tabular}{ccccccc} 
\toprule
Hard Negative Type    & R@100 & N@100 & R@500 & N@500 & R@1000 & N@1000  \\
\midrule
Baseline & 0.461$\pm$0.085 & 0.099$\pm$0.085 & 1.593$\pm$0.229 & 0.241$\pm$0.045 & 2.685$\pm$0.186 & 0.356$\pm$0.040 \\ 
Prerank\_neg & 0.575$\pm$0.095 & 0.127$\pm$0.028 & 1.883$\pm$0.170 & 0.291$\pm$0.030 & \textbf{2.931}$\pm$\textbf{0.142} & 0.401$\pm$0.030 \\ 
Coarse\_neg &  0.555$\pm$0.066 & 0.121$\pm$0.021 & 1.729$\pm$0.152 & 0.267$\pm$0.033 & 2.758$\pm$0.169 & 0.376$\pm$0.035 \\ 
Rank\_neg & 0.462$\pm$0.126 & 0.094$\pm$0.030 & 1.695$\pm$0.230 & 0.249$\pm$0.043 & 2.733$\pm$0.221 & 0.359$\pm$0.042 \\ 
Rank\_pos & 0.648$\pm$0.074 & 0.134$\pm$0.017 & 1.794$\pm$0.187 & 0.277$\pm$0.028 & 2.737$\pm$0.173 & 0.376$\pm$0.025 \\ 
Rerank\_neg & 0.577$\pm$0.091 & 0.119$\pm$0.019 & 1.804$\pm$0.208 & 0.274$\pm$0.034 & 2.724$\pm$0.242 & 0.371$\pm$0.036 \\ 
Rerank\_pos & \textbf{0.687}$\pm$\textbf{0.087} & \textbf{0.144}$\pm$\textbf{0.018} & \textbf{1.889}$\pm$\textbf{0.108} & \textbf{0.295}$\pm$\textbf{0.021} & 2.892$\pm$0.105 & 0.401$\pm$0.020 \\ 
Exposure\_neg & 0.603$\pm$0.093 & 0.137$\pm$0.016 & 1.860$\pm$0.207 & \textbf{0.295}$\pm$\textbf{0.032} & 2.902$\pm$0.221 & \textbf{0.405}$\pm$\textbf{0.033} \\ 
\bottomrule
\end{tabular}
\end{table}

\subsubsection{Interplay between Retrieval and Subsequent Stages}

The most important characteristic of industrial RS is the multi-stage. Every stage has its duty and mature paradigm. The goal of each stage is consistent, which is to fit the user's preference. Although models of all stages aim at fitting the user's preference, they can not capture the user's preference perfectly. Few people focus on the interplay between stages. The academic researchers lack available datasets and the industrial engineers only devote effort to the assigned stage. \citep{zheng2024full} has pointed out that there are two factors influencing the video's exposure and the user's feedback. First, it is the user's preference on the video. Second, it is the preference of the subsequent stage towards the video. For example, one video that the user likes is retrieved during the retrieving stage but is filtered out by the ranking model due to its imperfect preference estimation ability. This video is inefficient for the whole RS because it can not be exposed to the user at all. The optimal solution for the model of each stage is to select videos that satisfy the preference of the user and subsequent stages simultaneously. FS-LTR~\citep{zheng2024full} has proposed the Generalized Probability Ranking Principle (GPRP) to prove the solution proposed above is optimal theoretically. We implement FS-LTR in this section to see its effectiveness. The user's preference can be learned from the positive feedback samples and randomly sampling negative samples. In order to learn the preference of subsequent stages, we introduce additional ranking loss which forces the logits of samples from high-priority stages to be bigger than the logits of samples from low-priority stages. The priority of stages are \{positive:6, exposure\_neg:5, rerank\_pos:4, rank\_pos:4, rerank\_neg:3, rank\_neg:3, corase\_neg:2, pre-rank\_neg:1, random\_neg:0\}. Exposure\_neg represents the video that has been exposed to the user (realshow=1) but obtains negative feedback. This definition of priority applies throughout the paper. We always keep one positive sample with $200$ negative samples. We first introduce the stage preference one stage once by replacing random negatives with stage samples with BPR loss as Eq(~\ref{eq:recall_full_rank_loss}):
\begin{equation}
\label{eq:recall_full_rank_loss}
    L_{\textit{FS-LTR}}=\sum_{i=1}^N\sum_{j\in\{k:p_k<p_i\}}BPR(o_i,o_j)
\end{equation}
where $N$ equals $200$, $p_{i(k)}$ represents the priority level of sample $i(k)$, and $p_k<p_i$ means the priority level of sample $k$ is lower than sample $i$.

\begin{table}
\setlength{\tabcolsep}{2.5pt}
\renewcommand{\arraystretch}{1.0} 
\centering
\small
\caption{Recall(R) and NDCG(N) results (mean $\pm$ std) obtained by using a single different stage sample as the cascade sample during the retrieval stage, with units of \%. The best and baseline results are based on the paired $t$-test at the significance level $5\%$.}
\label{tab:retrieval_con1}
\begin{tabular}{ccccccc} 
\toprule
Cascade Type  & R@100 & N@100 & R@500 & N@500 & R@1000 & N@1000  \\
\midrule
Baseline & 0.461$\pm$0.085 & 0.099$\pm$0.085 & 1.593$\pm$0.229 & 0.241$\pm$0.045 & 2.685$\pm$0.186 & 0.356$\pm$0.040 \\ 
Prerank\_neg & 0.677$\pm$0.061 & 0.167$\pm$0.041 & 2.268$\pm$0.129 & 0.367$\pm$0.048 & 3.446$\pm$0.111 & 0.492$\pm$0.042 \\ 
Coarse\_neg & 0.665$\pm$0.120 & 0.163$\pm$0.045 & 2.253$\pm$0.052 & 0.361$\pm$0.037 & 3.371$\pm$0.090 & 0.479$\pm$0.038 \\ 
Rank\_neg & 0.704$\pm$0.150 & 0.173$\pm$0.049 & 2.282$\pm$0.250 & 0.373$\pm$0.055 & 3.410$\pm$0.203 & 0.491$\pm$0.052 \\ 
Rank\_pos & 0.685$\pm$0.094 & 0.151$\pm$0.025 & 2.191$\pm$0.085 & 0.340$\pm$0.023 & 3.346$\pm$0.078 & 0.462$\pm$0.019 \\ 
Rerank\_neg & 0.707$\pm$0.083 & 0.163$\pm$0.024 & 2.273$\pm$0.121 & 0.359$\pm$0.024 & 3.338$\pm$0.083 & 0.471$\pm$0.022 \\ 
Rerank\_pos & 0.795$\pm$0.108 & 0.176$\pm$0.025 & 2.263$\pm$0.078 & 0.361$\pm$0.017 & 3.394$\pm$0.048 & 0.480$\pm$0.016 \\ 
Exposure\_neg & 0.692$\pm$0.071 & 0.156$\pm$0.028 & 2.150$\pm$0.108 & 0.340$\pm$0.033 & 3.266$\pm$0.183 & 0.458$\pm$0.036 \\ 
FS-LTR & \textbf{0.803}$\pm$\textbf{0.095} & \textbf{0.215}$\pm$\textbf{0.027} & \textbf{2.466}$\pm$\textbf{0.090} & \textbf{0.425}$\pm$\textbf{0.029} & \textbf{3.606}$\pm$\textbf{0.060} & \textbf{0.545}$\pm$\textbf{0.024} \\
\bottomrule
\end{tabular}
\end{table}

We have the following findings from experiment results in Table~\ref{tab:retrieval_con1}. (1) FS-LTR gains performance enhancement when introducing the sample of each stage respectively compared to the baseline. (2) Under the same negative setting, FS-LTR can achieve better results compared with the results of hard negative mining in Table~\ref{tab:retrieval_hard}. (3) As the K of Recall@K becomes smaller, the performance improvement becomes better. For example, when we add $1$ \textit{pre-rank\_neg} as hard negative, the relative promotion of Recall@100, 500, 100 are $46.8\%$, $42.4\%$, $28.3\%$ respectively. NDCG@K holds the same trends. We also try to introduce multiple samples from more stages gradually to investigate the effectiveness of modeling more subsequent stages' preferences in subsection~\ref{retrieval_interplay_exp} of Appendix.

\subsection{Coarse Ranking}
Coarse ranking receives favorable videos from the pre-ranking stage and filters less favorable videos to fulfill its duty. As the candidate videos are more similar and not easy to distinguish, coarse ranking models take more feature fields as input and use a more complex neural network to ensure their modeling capacity. However, there are $3,000$ videos to be scored in our scenario, the two-tower structure is still the best choice. We take DSSM~\citep{huang2013learning} as the coarse ranking model. The Multi-layer Perceptron (MLP) of the user and video towers in DSSM are set to be [128, 64, 32]. Existing coarse ranking models are almost learned on the exposure of positive and negative samples. AUC (Area Under the Curve) on the testing exposure samples is employed to assess the algorithm's performance. \textit{Effective\_view} is the learning signal. Following the retrieval experiment, data from the first $36$ days is for training and the last day's data is for evaluation. The feature fields include \textit{user\_id, device\_id, age, gender, province, video\_id, author\_id, category\_level\_one, category\_level\_two, upload\_type, upload\_timestamp, request\_timestamp}. The \textit{upload\_timestamp} and  \textit{request\_timestamp} are divided into the \textit{week, day, hour} feature fields. Besides, we add the user's past $50$ effective\_videos as the behavior sequence. We process effective behavior sequences through mean pooling. We apply embedding for all the feature fields and set the embedding dimension to $8$. The batch size is $1,024$ and the learning rate is $1e-2$. Binary Cross Entropy is the loss and Adam is used for optimization. We also utilize the stage samples to explore the auxiliary ranking task and user behavior sequence modeling in the coarse ranking model, both of which boost the AUC metric greatly. The methods together with the experiment results and analysis are in subsections~\ref{coarse_auxiliary_ranking} and \ref{coarse_behavior_modeling} of Appendix. 

\subsubsection{Data Distribution Shift}
Data distribution shift is a longstanding problem in RS. Due to the absence of datasets containing stage samples, few works~\citep{ma2018entire,qin2022rankflow} focus on the problem in the coarse ranking stage. The coarse ranking model is trained based on the exposed samples which contains $6$ videos at most but has to score $3,000$ videos in each request. The data distribution between training and testing exists huge inconsistency. What's worse, the AUC metric evaluated on the exposure space for guiding the offline algorithm's optimization is inconsistent with the online scenario~\citep{song2022rethinking,zhang2023rethinking}. The collected stage samples make the evaluation space more consistent with the online situation. Following ~\citep{zhang2023rethinking}, we apply the Recall@K metric which is consistent with the effect of online business. Because we saved all the videos in the ranking stage on February 18, 2024, the candidate set for calculating the Recall@K and NDCG@K is composed of the videos in the ranking stage together with videos of coarse\_neg. We set K to $100, 200$. We also report the classical AUC metric. We try to directly supplement the stage samples as extra negative samples into the training data. Although it's possible to introduce false negative videos, this still can reduce the difference in data distribution between training and testing largely. However, supplementing extra negative samples increases the machine overload. Thus, we show the relationship between the performance and the quantity of the additional negative samples.

We have the following conclusions from the result in Table \ref{tab:Coarse_hard}. 
(1) Supplementing stage videos as extra negative samples can largely enhance the Recall and NDCG metric. The improvement can be attributed to the consistency of data distribution between training and testing. 
(2) When increasing the quantity of extra negative samples, the improvement becomes greater. And introducing partial or all videos from all corresponding stages gains the best results. This indicates the more consistent data distribution between training and testing, the more improvement. 
(3) Introducing videos from \textit{rank\_pos} and \textit{rerank\_pos} gains light enhancement compared to \textit{coarse/rank/rerank\_neg}. We think that there are false negative samples that mislead the model's learning. 
(4) The classical AUC metric has opposite trends. After adding extra negative samples, the gap between the training data distribution and the data distribution of exposure space for evaluating AUC enlarges. As we mentioned in the retrieval section, there exist hardness level among different stage samples. Expanding negatives degrades the model's ability to distinguish hard negatives (exposed un-effective\_view samples) but enhances the capability of recognizing less hard negatives (videos from stages). 

\begin{table}
\setlength{\tabcolsep}{3pt}
\renewcommand{\arraystretch}{1.0} 
\centering
\small
\caption{The result (mean $\pm$ std) of using different stages' samples as extra negatives for Coarse Ranking. The best and baseline results are based on the paired $t$-test at the significance level $5\%$.}
\label{tab:Coarse_hard}
\begin{tabular}{c|c|cc|cccc} 
\toprule
Neg Type & \#N & AUC & LogLoss & Recall@100 & NDCG@100 & Recall@200 & NDCG@200 \\
\midrule
Baseline & - & \textbf{0.718}$\pm$\textbf{0.001} & \textbf{0.592}$\pm$\textbf{0.003} & 0.271$\pm$0.027 & 0.059$\pm$0.027 & 0.535$\pm$0.009 & 0.096$\pm$0.003 \\ 
\midrule
\multirow{2}{*}{Coarse\_neg} & 1 & 0.705$\pm$0.002 & 0.608$\pm$0.006 & 0.321$\pm$0.012 & 0.072$\pm$0.027 & 0.597$\pm$0.038 & 0.111$\pm$0.003 \\ 
&10 & 0.633$\pm$0.016 & 0.773$\pm$0.018 & 0.392$\pm$0.012 & 0.088$\pm$0.004 & 0.668$\pm$0.007 & 0.126$\pm$0.003 \\ 
\midrule
\multirow{2}{*}{Rank\_neg} & 1 & 0.704$\pm$0.002 & 0.615$\pm$0.004 & 0.353$\pm$0.013 & 0.079$\pm$0.003 & 0.638$\pm$0.004 & 0.118$\pm$0.002 \\
&10 & 0.618$\pm$0.016 & 0.825$\pm$0.027 & 0.454$\pm$0.011 & 0.102$\pm$0.005 & 0.726$\pm$0.005 & 0.140$\pm$0.004 \\
\midrule
\multirow{2}{*}{Rank\_pos} & 1 & 0.704$\pm$0.001 & 0.603$\pm$0.001 & 0.275$\pm$0.027 & 0.061$\pm$0.004 & 0.557$\pm$0.005 & 0.100$\pm$0.001 \\
&10 & 0.623$\pm$0.020 & 0.769$\pm$0.003 & 0.290$\pm$0.002 & 0.069$\pm$0.004 & 0.591$\pm$0.019 & 0.111$\pm$0.004 \\
\midrule
\multirow{2}{*}{Rerank\_neg} & 1 & 0.702$\pm$0.001 & 0.616$\pm$0.005 & 0.337$\pm$0.007 & 0.076$\pm$0.001 & 0.605$\pm$0.002 & 0.113$\pm$0.001 \\ 
&10 & 0.608$\pm$0.021 & 0.821$\pm$0.019 & 0.380$\pm$0.014 & 0.084$\pm$0.003 & 0.673$\pm$0.004 & 0.125$\pm$0.003 \\ 
\midrule
\multirow{2}{*}{Rerank\_pos} & 1 & 0.703$\pm$0.001 & 0.607$\pm$0.003 & 0.264$\pm$0.011 & 0.060$\pm$0.003 & 0.548$\pm$0.003 & 0.099$\pm$0.002 \\ 
&10 & 0.618$\pm$0.024 & 0.782$\pm$0.025 & 0.285$\pm$0.015 & 0.069$\pm$0.003 & 0.587$\pm$0.011 & 0.111$\pm$0.003 \\ 
\midrule
\multirow{2}{*}{All} & 1 & 0.662$\pm$0.006 & 0.704$\pm$0.011 & 0.386$\pm$0.010 & 0.084$\pm$0.001 & 0.676$\pm$0.008 & 0.125$\pm$0.001 \\
&10 & 0.563$\pm$0.004 & 1.243$\pm$0.030 & \textbf{0.455}$\pm$\textbf{0.004} & \textbf{0.105}$\pm$\textbf{0.001} & \textbf{0.728}$\pm$\textbf{0.004} & \textbf{0.144}$\pm$\textbf{0.001} \\ 
\bottomrule
\end{tabular}
\end{table}

\subsubsection{Interplay between Coarse Ranking and Subsequent Stages}
\begin{table}
\setlength{\tabcolsep}{3pt}
\renewcommand{\arraystretch}{1.0} 
\centering
\small
\caption{The result (mean $\pm$ std) of interplay between Coarse Ranking and Subsequent Stages. The best and baseline results are based on the paired $t$-test at the significance level $5\%$.}
\label{tab:coarse_interplay}
\begin{tabular}{c|cc|cccc} 
\toprule
Method & AUC & LogLoss & Recall@100 & NDCG@100 & Recall@200 & NDCG@200 \\
\midrule
Baseline  & \textbf{0.718}$\pm$\textbf{0.001} & \textbf{0.592}$\pm$\textbf{0.003} & 0.271$\pm$0.027 & 0.059$\pm$0.027 & 0.535$\pm$0.009 & 0.096$\pm$(0.003 \\ 
PositiveRank  & 0.554$\pm$0.005 & 1.040$\pm$0.051 & 0.457$\pm$0.001 & 0.112$\pm$0.001 & 0.723$\pm$0.002 & 0.149$\pm$0.001 \\  
% Plain-FullRank  & 0.470 & 1.064 & 0.236 & 0.072 & 0.397 & 0.098 & 0.659 & 0.135 \\ 
FS-LTR  & 0.473$\pm$0.013 & 1.253$\pm$0.071 & \textbf{0.475}$\pm$\textbf{0.002} & \textbf{0.119}$\pm$0.001 & \textbf{0.734}$\pm$\textbf{0.002} & \textbf{0.155}$\pm$\textbf{0.001} \\ 
\bottomrule
\end{tabular}
\end{table}

FS-LTR is a general principle and is applicable in the coarse ranking stage. We implement FS-LTR with samples of \textit{positive, exposure\_neg, rerank\_pos, rerank\_neg, rank\_pos, rank\_neg, coarse\_neg}, which is the inference space of the coarse ranking model. In order to apply the loss~\ref{eq:recall_full_rank_loss}, we aggregate samples of the same request into the same batch. We also add a contrast experiment PostiveRank, in which we just make the logits of positive samples bigger than the logits of the other samples. The result in Table~\ref{tab:coarse_interplay} shows that FS-LTR can achieve the best performance on the Recall/NDCG. It demonstrates the necessity of learning the preferences of both the user and the subsequent stages. 

\subsection{Ranking}
Ranking is nearly the most important stage in the industrial multi-stage RS and has been studied sufficiently. It determines the displayed items to the user. Its candidate video set is the output of the coarse ranking stage. Given the importance and difficulty of the task, ranking model has the most complex neural network structure and uses most feature fields. The time cost is acceptable because it only needs to score $500$ videos. We utilize DIN~\citep{zhou2018deep} as the ranking model. The architecture of DIN's MLP is [128, 128, 32, 1]. Ranking model is also learned on the exposure space and evaluates AUC on testing exposure samples. For the experiment settings, the ranking model remains the same as the coarse ranking model. We also use the stage samples to explore the auxiliary ranking task and user behavior sequence modeling in the ranking model, both of which improve the classical AUC greatly. Detail of methods and experiments are in subsection~\ref{ranking_auxiliary_ranking} and~\ref{ranking_behavior_modeling} of Appendix. 

\subsubsection{Data Distribution Shift}
\begin{table}
\setlength{\tabcolsep}{3pt}
\renewcommand{\arraystretch}{1.0} 
\centering
\small
\caption{The result (mean $\pm$ std) of using different stages' samples as extra negatives for Ranking. The best and baseline results are based on the paired $t$-test at the significance level $5\%$.}
\label{tab:rank_hard}
\begin{tabular}{c|c|cc|cccccc} 
\toprule

Neg Type & \#N & AUC & LogLoss & Recall@50 & NDCG@50 & Recall@100 & NDCG@100 &\\
\midrule
Baseline & - & \textbf{0.727}$\pm$\textbf{0.001} & \textbf{0.583}$\pm$\textbf{0.003} & 0.169$\pm$0.005 & 0.045$\pm$0.002 & 0.319$\pm$0.008 & 0.069$\pm$0.002 \\ 
\midrule
\multirow{2}{*}{Rank\_neg} & 1 & 0.711$\pm$0.001 & 0.610$\pm$0.008 & 0.223$\pm$0.005 & 0.061$\pm$0.002 & 0.395$\pm$0.007 & 0.088$\pm$0.002 \\
&10 & 0.645$\pm$0.003 & 0.810$\pm$0.032 & 0.264$\pm$ 0.012& 0.074$\pm$0.004 & 0.454$\pm$0.014 & 0.105$\pm$0.005 \\ 
\midrule
\multirow{2}{*}{Rank\_pos} & 1 & 0.711$\pm$0.001 & 0.604$\pm$0.008 & 0.176$\pm$0.005 & 0.047$\pm$0.001 & 0.327$\pm$0.010 & 0.072$\pm$0.002 \\ 
&10 & 0.653$\pm$0.002 & 0.724$\pm$0.029 & 0.185$\pm$0.009 & 0.049$\pm$0.003 & 0.331$\pm$0.015 & 0.073$\pm$0.003 \\ 
\midrule
\multirow{2}{*}{Rerank\_neg} & 1 & 0.708$\pm$0.001 & 0.616$\pm$0.006 & 0.215$\pm$0.003 & 0.059$\pm$0.001 & 0.380$\pm$0.006 & 0.085$\pm$0.001 \\ 
&10 & 0.624$\pm$0.005 & 0.815$\pm$0.028 & 0.232$\pm$0.018 & 0.064$\pm$0.005 & 0.406$\pm$0.031 & 0.092$\pm$0.007 \\ 
\midrule
\multirow{2}{*}{Rerank\_pos} & 1 & 0.711$\pm$0.002 & 0.608$\pm$0.006 & 0.170$\pm$0.012 & 0.045$\pm$0.003 & 0.319$\pm$0.019 & 0.069$\pm$0.004 \\ 
&10 & 0.646$\pm$0.002 & 0.782$\pm$0.033 & 0.183$\pm$0.016 & 0.048$\pm$0.005 & 0.335$\pm$0.009 & 0.073$\pm$0.003 \\ 
\midrule
\multirow{2}{*}{All} & 1 & 0.675$\pm$0.003 & 0.697$\pm$0.010 & 0.234$\pm$0.005 & 0.064$\pm$0.002 & 0.411$\pm$0.004 & 0.093$\pm$0.002 \\ 
&10 & 0.602$\pm$0.005 & 1.076$\pm$0.049 & \textbf{0.278}$\pm$\textbf{0.027} & \textbf{0.078}$\pm$\textbf{0.007} & \textbf{0.467}$\pm$\textbf{0.038} & \textbf{0.108}$\pm$\textbf{0.009} \\ 
\bottomrule
\end{tabular}
\end{table}

The ranking model also suffers from the data distribution shift problem. In each request, there are at most $6$ exposure samples for training but $500$ videos to be scored. The data distribution gap between training and testing still exists. Fortunately, the inconsistency is not as serious as the coarse ranking model. The exploration experiment setting for alleviating the data distribution shift problem is the same as the coarse ranking model including motivation, method, and evaluation metrics. The difference is that samples of \textit{coarse\_neg} are excluded for training and evaluation because they are not in the ranking model's candidate video set. The result is shown in Table \ref{tab:rank_hard}. We can find that the more consistent the data distribution between training and testing, the Recall and NDCG gain more improvement. Other conclusions are the same as coarse ranking and we don't repeat them here. 

\subsubsection{Interplay between Ranking and Subsequent Stages}
\begin{table}
\setlength{\tabcolsep}{5pt}
\renewcommand{\arraystretch}{1.0} 
\centering
\small
\caption{The result (mean $\pm$ std) of interplay between Ranking and Subsequent Stages. The best and baseline results are based on the paired $t$-test at the significance level $5\%$.}
\label{tab:rank_interplay}
\begin{tabular}{c|cc|cccccc} 
\toprule

Method & AUC & LogLoss & R@50 & N@50 & R@100 & N@100 \\
\midrule
Baseline & \textbf{0.727}$\pm$\textbf{0.001} & \textbf{0.583}$\pm$\textbf{0.003} & 0.169$\pm$0.005 & 0.045$\pm$0.002 & 0.319$\pm$0.008 & 0.069$\pm$0.002 \\  
PositiveRank  & 0.564$\pm$0.003 & 1.466$\pm$0.313 & 0.309$\pm$0.016 & 0.093$\pm$0.006 & 0.506$\pm$0.014 & 0.125$\pm$0.006 \\  
% Plain-FullRank  & 0.456 & 0.954 & 0.271 & 0.083 & 0.452 & 0.112 & 0.733 & 0.151 \\ 
FS-LTR  & 0.461$\pm$0.005 & 1.215$\pm$0.391 & \textbf{0.323}$\pm$\textbf{0.012} & \textbf{0.098}$\pm$\textbf{0.003} & \textbf{0.525}$\pm$\textbf{0.014} & \textbf{0.131}$\pm$\textbf{0.004} \\ 
\bottomrule
\end{tabular}
\end{table}

We also conduct FS-LTR in the ranking stage. The experiment settings are mostly the same as coarse ranking except that training samples are from \textit{positive, exposure\_neg, rank\_neg, rank\_pos, rerank\_neg, rerank\_pos}, which is the inference space of the ranking model. PositiveRank serves as the contrast purpose. The results are summarized in the Table~\ref{tab:rank_interplay} and conclusions are the same as coarse ranking.

\section{Limitations}
RecFlow, while valuable in lots of recommendation research problems, also has its own drawbacks. Understanding advantages and disadvantages is vital for ensuring accurate academic use. First, we collect data from only one recommendation scenario which causes RecFlow can not be applied to the multi/cross-domain recommendation. Second, RecFlow can't advance the research of multimodal recommendation because of lacking multimodal features such as text and image. On the other hand, it needs more hardware resource cost because RecFlow contains 1,924,337,704 instances.

\section{Conclusions}
In this paper, we propose a new dataset called RecFlow. Unlike all previously published recommendation datasets, RecFlow captures information across the entire pipeline of an industrial recommendation system. We believe this is highly valuable as it will provide researchers with unprecedented convenience for studying multi-stage recommendations. 
We also conduct extensive preliminary experiments using RecFlow in retrieval, coarse ranking, and ranking stages. The experimental results demonstrate that utilizing stage samples indeed enhances recommendations.
% We also conduct extensive preliminary experiments using RecFlow, including hard negative mining, interplay between stages, auxiliary ranking task, and modeling competitive relationships between stage samples. The experimental results demonstrate that utilizing stage samples indeed enhances recommendations.

% \clearpage

\bibliography{iclr2025_conference}
\bibliographystyle{iclr2025_conference}

\clearpage

\appendix
\section{Appendix}
\subsection{Feature Description} \label{feature_description}
The \textit{request\_id} identifies each recommendation request and \textit{request\_timestamp} represents the time when the recommendation request arises. Every user has a unique ID named \textit{user\_id}. \textit{device\_id} means the device that initiates the recommendation request. We also provide the user's profile information including \textit{age, gender, province}. \textit{Age} is grouped into ten buckets. \textit{video\_id} identifies each video. \textit{author\_id} represents the one who uploads the video. We also record the video's attributes involving \textit{category\_level\_one, category\_level\_two, upload\_type, upload\_timestamp, duration}. \textit{category\_level\_one, category\_level\_two} are categories of the video, where \textit{category\_level\_one} is the coarse-grained category (e.g. sports, history, k-pop, etc.) and \textit{category\_level\_two} indicates the fine-grained category (e.g. UEFA Champions League, Ming Dynasty, BLACKPINK, etc). The \textit{upload\_type} and \textit{upload\_timestamp} stand for the type of the video (e.g. micro-video, long-video, picture, etc) and the time when the video was uploaded. \textit{duration} is the video's lasting time. Next, we describe the fields identifying the stage information.  The \textit{effective\_view} and \textit{long\_view} are the binary features (0 and 1) defined according to business interest. \textit{long\_view} is more strict than \textit{effective\_view}. \textit{like} indicates whether the user clicks the \textcolor{red}{$\heartsuit$} button. \textit{follow} means the user follows the video's author. \textit{forward} represents the user sharing the video. \textit{comment} stands for whether the user makes some text review about the video. Note that the feedback values of the unexposed video are all set to 0. The fields of \textit{request\_id, user\_id, device\_id, age, gender, province, video\_id, author\_id, category\_level\_one, category\_level\_two, upload\_type} are all have been anonymized ensuring the privacy protection.

\subsection{Dataset Comparison} \label{dataset_comparison}

\begin{table}[h]
\caption{The characteristic comparison of different recommendation datasets.}
\label{tab:dataset_comparison}
\centering
\small
\setlength{\tabcolsep}{3.6pt}
\renewcommand{\arraystretch}{1.0} 
\begin{tabulary}{\textwidth}{LCCCCCC} 
\toprule
Dataset                 & Stage\ Sample & Type\_feedbacks & \#Users & \#Interactions & True\_neg & Req\_id  \\
\midrule
MovieLens-20M~\citep{harper2015movielens}           &  \ding{55}              & 1           & 138K    & 20M            &   \ding{55}               &  \ding{55}               \\
Amazon~\citep{ni2019justifying}                  &\ding{55}             & 2           & /       & 233M           & \ding{55}                 &   \ding{55}              \\
Yelp~\citep{asghar2016yelp}                    & \ding{55}               & 1           & 1.9M    & 8M             &  \ding{55}                &  \ding{51}               \\
Taobao~\citep{zhu2018learning}                  &  \ding{55}              & 4           & 987K    & 100M           &   \ding{55}               &   \ding{55}              \\ 
% Ali\_Display\_Ad\_Click &    \ding{55}            & 4           & 1.1M    & 26M            &   \ding{51}               &    \ding{55}         \\
% Tmall                   &   \ding{55}             & 2           & 963K    & 44M            &   \ding{51}               &    \ding{55}             \\
TenRec-QKV~\citep{yuan2022tenrec}                 &   \ding{55}             & 4           & 5.0M    & 142M           &   \ding{51}               &     \ding{55}            \\
TenRec-QKA~\citep{yuan2022tenrec}              &    \ding{55}            & 6           & 1.3M    & 46M            &   \ding{55}               &       \ding{55}          \\
KuaiRec~\citep{gao2022kuairec}                 &    \ding{55}            & 1           & 7K      & 12M            &   \ding{51}               &         \ding{55}        \\
KuaiRand~\citep{gao2022kuairand}                &    \ding{55}            & 6           & 27K     & 322M           &    \ding{51}              &       \ding{55}          \\
KuaiSAR~\citep{sun2023kuaisar}                 &     \ding{55}           & 9           & 26K     & 19M            &    \ding{51}              &      \ding{55}           \\
\midrule
RecFlow                 &    \ding{51}(1.9B)            & 7           &   42K      &     38M           &   \ding{51}               &      \ding{51}           \\
\bottomrule
\end{tabulary}
\end{table}

\subsection{Retrieval: The effect of the number of hard negatives in retrieval stage} \label{retrieval_num_hard_neg}
\begin{table}
\setlength{\tabcolsep}{3.6pt}
\renewcommand{\arraystretch}{1.0} 
\centering
\caption{Recall(R) and NDCG(N) results obtained by using 2 or 10 different stage sample as the hard negative sample during the retrieval stage, with units of \%.}
\label{tab:retrieval_hardn}
\begin{tabular}{cccccccc} 
\toprule
HN Type & \#HN   & R@100 & N@100 & R@500 & N@500 & R@1000 & N@1000  \\
\midrule
Baseline & - & 0.461 & 0.099 & 1.593 & 0.241 & 2.685 & 0.356 \\ 
\midrule
\multirow{2}{*}{Prerank\_neg} & 2 & 0.474 & 0.108 & 1.664 & 0.257 & 2.574 & 0.352 \\ 
&10 & 0.457 & 0.101 & 1.515 & 0.236 & 2.319 & 0.321 \\ 
\midrule
\multirow{2}{*}{Coarse\_neg} & 2 & \textbf{0.634} & 0.140 & \textbf{1.948} & \textbf{0.305} & 2.844 & 0.400 \\ 
&10 & 0.524 & 0.125 & 1.564 & 0.256 & 2.347 & 0.339 \\ 
\midrule
\multirow{2}{*}{Rank\_neg} & 2 & 0.492 & 0.104 & 1.687 & 0.254 & 2.645 & 0.355 \\ 
&10 & 0.323 & 0.069 & 1.321 & 0.193 & 2.231 & 0.289 \\ 
\midrule
\multirow{2}{*}{Rank\_pos} & 2 & 0.589 & 0.126 & 1.846 & 0.284 & 2.722 & 0.376 \\ 
&10 & 0.544 & 0.109 & 1.544 & 0.235 & 2.310 & 0.315 \\ 
\midrule
\multirow{2}{*}{Rerank\_neg} & 2 & 0.606 & 0.120 & 1.849 & 0.277 & 2.831 & 0.381 \\ 
&10 & 0.336 & 0.070 & 1.186 & 0.176 & 2.032 & 0.265 \\ 
\midrule
\multirow{2}{*}{Rerank\_pos} & 2 & 0.428 & 0.091 & 1.584 & 0.236 & 2.551 & 0.338 \\ 
&10 & 0.219 & 0.043 & 0.954 & 0.135 & 1.819 & 0.226 \\  
\midrule
\multirow{2}{*}{exposure\_neg} & 2 & 0.576 & 0.138 & 1.854 & 0.300 & \textbf{2.866} & \textbf{0.407} \\ 
&10 & 0.629 & \textbf{0.142} & 1.924 & \textbf{0.305} & 2.856 & 0.403 \\ 
\bottomrule
\end{tabular}
\end{table}

The result in Table \ref{tab:retrieval_hardn} is the result of varying the number of hard negative samples from each stage. We set the number to $2$ and $10$ for observation. (1) Increasing the number of hard negative videos from \textit{prerank\_neg} can further improve the performance but with diminishing marginal effect. (2) For \textit{coarse\_neg, rank\_pos, rerank\_neg, rerank\_pos, exposure\_neg}, adding videos from them as hard negative samples degrades the performance. The closer the stage to the positive feedback, the more degradation. The phenomenon demonstrates that there exists hardness level between videos from different stages. The closer to the positive feedback the stage, the videos it contains are more difficult for the retrieval model to distinguish. (3) When we increase the number of videos from \textit{rank\_neg} from $1$ to $2$, the performance has somewhat boosted. However, it still suffers from a severe performance drop when taking $10$ \textit{rank\_neg} hard negative videos. As pointed out in~\citep{he2014practical,zhang2023divide}, the ratio between easy and hard negatives has a critical influence on the performance. We guess that harder negatives need more easy negatives and leave the hardness and ratio of hard negative samples for further research.

\subsection{Retrieval: More Result of Retrieval's Interplay Experiment} \label{retrieval_interplay_exp}

\begin{table}
\setlength{\tabcolsep}{4pt}
\renewcommand{\arraystretch}{1.0} 
\centering
\small
\caption{Recall(R) and NDCG(N) results obtained by using different combinations of stage samples as cascade samples during the retrieval stage, with units of \%. CN-PN represents the use of coarse\_neg and preran\_neg. R-CN-PN represents the use of exposure\_neg, coarse\_neg, and prerank\_neg. ALL represents the use of all stage samples in the current request.}
\label{tab:retrieval_con2}
\begin{tabular}{ccccccc} 
\toprule
Cascade Type & R@100 & N@100 & R@500 & N@500 & R@1000 & N@1000  \\
\midrule
Baseline & 0.461 & 0.099 & 1.593 & 0.241 & 2.685 & 0.356 \\  
CN-PN & \textbf{0.803} & \textbf{0.215} & 2.466 & \textbf{0.425} & 3.606 & \textbf{0.545} \\ 
EN-CN-PN & 0.771 & 0.198 & \textbf{2.481} & 0.413 & \textbf{3.648 }& 0.536 \\ 
All & 0.663 & 0.150 & 1.972 & 0.315 & 3.018 & 0.425 \\ 
\bottomrule
\end{tabular}
\end{table}

The result of Table~\ref{tab:retrieval_con2} is the experiment of introducing samples from more different stages into the FS-LTR. We can find that directly introducing samples from all stages is better than the baseline but is not the best. The setting of CN-PN achieves the best in our exploration, which indicates that ranking regularization between some priority levels may be unnecessary. We leave the in-depth exploration for future research. 

\subsection{Coarse Ranking: Auxiliary Ranking Task} \label{coarse_auxiliary_ranking}

\begin{table}
\setlength{\tabcolsep}{5pt}
\renewcommand{\arraystretch}{1.0} 
\centering
\small
\caption{The result of the auxiliary ranking task for the coarse ranking stage.}
\label{tab:coarse_aux}
\begin{tabular}{c|cc|cccccc} 
\toprule

Method & AUC & LogLoss & R@100 & N@100 & R@200 & N@200 \\
\midrule
Baseline  & 0.718 & 0.592 & 0.271 & 0.059 & 0.535 & \textbf{0.096} \\ 
\midrule
w/ AuxLoss  & \textbf{0.721} & \textbf{0.588} & \textbf{0.287} & \textbf{0.061} & \textbf{0.541} & \textbf{0.096} \\ 
\bottomrule
\end{tabular}
\end{table}

Increasing the ranking ability of the model trained with pointwise loss function (e.g. Click-through Rate prediction model)  by adding an auxiliary ranking task has gained much attention recently~\citep{yan2022scale,bai2023regression,sheng2023joint,liu2024at4ctr,lin2024understanding}. The auxiliary ranking task forces the logits of positive samples to be bigger than negative samples within the same batch or session through pairwise or listwise ranking loss. Inspired by these works, we propose a new auxiliary ranking task by forcing the logits of positive samples bigger than the stage samples of the same request. There is no ranking regularization on the negative samples. Note that the stage samples are only for auxiliary loss. The total loss function is as Eq(~\ref{eq:coarse_rank_loss}).
\begin{equation}
    \label{eq:coarse_rank_loss}
    L=\frac{1}{N}\sum_{i=1}^NBCEWithLogit(o_i,y_i)+\alpha\ast\frac{1}{N_+K}\sum_{j=1}^{N_+}\sum_{j_k=1}^{K}BPR(o_j,o_{j_k})
\end{equation}
where $N$ is the batch size, $N_+$ is the number of positive samples in the batch, $j_k$ represents the stage sample within the same request as $j$, $K$ is the size of stage samples, and $o$ is the logit output by DSSM. $\alpha$ is the weight of auxiliary ranking loss. In the experiment, we use all the stage samples from \textit{coarse\_neg, rank\_neg, rank\_pos, rerank\_neg, rerank\_pos} stages. The result in Table~\ref{tab:coarse_aux} shows that the AUC increase by $0.002$ and the Logloss decrease by $0.004$, which is a significant improvement~\citep{guo2017deepfm}. Recall and NDCG also gain improvement. The designed auxiliary ranking task promotes both the classical and the newly proposed metrics, which demonstrates its effectiveness.

\subsection{Coarse Ranking:  User Behavior Sequence Modeling} \label{coarse_behavior_modeling}
Competitive relation modeling has been attracting attention in user behavior sequence modeling (UBM) recently~\citep{zheng2022implicit,hou2023deep,fan2022modeling,li2023decision}. Its motivation is that the user's feedback on items is also influenced by the displayed context. For example, if one user likes red T-shirts, he/she will click a pink T-shirt surrounded by items which he/she is not interested in, but he/she will click the red T-shirt surrounded by pink, blue, and yellow T-shirts. The competitive relation among displayed items has an impact on the user's feedback. There also exists competitive relation among the videos in the \textit{rerank/rank\_pos} stages. These videos compete for exposure to the user. Inspired by~\citep{hou2023deep}, we explore introducing the competitive information in the stage samples into the UBM. For the user's past effective\_view videos $S=[v_1,v_2,...,v_{50}]$, we regard $10$ videos in the \textit{rank\_pos} from the same request of each effective\_view video in $S$ as the competitive information. We represent the competitive relation as $C=[[v_{1,1},v_{1,2},...,v_{1,10}],[v_{2,1},v_{2,2},...,v_{2,10}],...,[v_{50,1},v_{50,2},...,v_{50,10}]]$. We apply the hierarchical attention algorithm to model the competitive relation. First, we perform target attention between each effective\_view behavior $v_i$ and its competing context $[v_{i,1},v_{i,2},...,v_{i,10}]$. We will obtain the refined competing behavior representation $E=[c_1,c_2,...,c_{50}]$. Then, we do mean pooling on $E$ to the user's competitive relation aware interest $competing\_interest$. Table \ref{tab:coarse_behav} shows the experiment results. Both AUC and Logloss are improved significantly by $0.004$. What's more, Recall@{100,200} and NDCG@{100,200} also get better performance. The video competitive information in the \textit{rank\_pos} is a useful signal for UBM. The result indicates there exists a method that can improve both the classical AUC/Logloss and the newly applied Recall/NDCG metric.
\begin{table}
\setlength{\tabcolsep}{5pt}
\renewcommand{\arraystretch}{1.0} 
\centering
\small
\caption{The result of competitive relation modeling in UBM during coarse ranking stage.}
\label{tab:coarse_behav}
\begin{tabular}{c|cc|cccc} 
\toprule

Method & AUC & LogLoss & R@100 & N@100 & R@200 & N@200 \\
\midrule
Baseline  & 0.718 & 0.592 & 0.271 & 0.059 & 0.535 & 0.096 \\ 
\midrule
Competing Seq  & \textbf{0.722} & \textbf{0.588} & \textbf{0.293} & \textbf{0.064} & \textbf{0.574} & \textbf{0.103} \\ 
\bottomrule
\end{tabular}
\end{table}

\subsection{Ranking: Auxiliary Ranking Task} \label{ranking_auxiliary_ranking}

We also conduct the auxiliary ranking task in the ranking stage. The ranking loss is still Eq(\ref{eq:coarse_rank_loss}). The stage samples used in the ranking loss come from \textit{rank\_neg, rank\_pos, rerank\_neg, rerank\_pos} stages. The results are summarized in the Table~\ref{tab:rank_aux}. We can draw findings the same with the auxiliary ranking task of coarse ranking, which verifies the broad effectiveness of the auxiliary ranking loss based on the stage samples. 

\begin{table}
\setlength{\tabcolsep}{5pt}
\renewcommand{\arraystretch}{1.0} 
\centering
\small
\caption{The result of the auxiliary ranking task for the ranking stage.}
\label{tab:rank_aux}
\begin{tabular}{c|cc|cccc} 
\toprule
Method & AUC & LogLoss & R@50 & N@50 & R@100 & N@100 \\
\midrule
Baseline & 0.727 & \textbf{0.583} & \textbf{0.169} & \textbf{0.045} & \textbf{0.319} & \textbf{0.069} \\ 
\midrule
w/ AuxLoss  & \textbf{0.729} & \textbf{0.583} & 0.168 & \textbf{0.045} & 0.316 & 0.068 \\ 
\bottomrule
\end{tabular}
\end{table}

\subsection{Ranking: User Behavior Sequence Modeling} \label{ranking_behavior_modeling}
\begin{table}
\setlength{\tabcolsep}{5pt}
\renewcommand{\arraystretch}{1.0} 
\centering
\small
\caption{The result of competitive relation modeling in UBM during ranking stage.}
\label{tab:rank_behav}
\begin{tabular}{c|cc|cccc} 
\toprule
& AUC & LogLoss & R@50 & N@50 & R@100 & N@100 \\
\midrule
Baseline & 0.727 & 0.583 & \textbf{0.169} & \textbf{0.045} & \textbf{0.319} & \textbf{0.069} \\ 
\midrule
Competing Seq  & \textbf{0.732} & \textbf{0.578} & 0.168 & \textbf{0.045} & 0.313 & 0.068 \\ 
\bottomrule
\end{tabular}
\end{table}

Competing modeling is also explored in the ranking stage. We still choose the videos in the \textit{rank\_pos} as the competing context. We change the modeling method and make it more suitable for DIN. After acquiring the refined competing behavior sequence representation $E=[c_1,c_2,...,c_{50}]$, we perform target attention between target video $v_{target}$ and $E$. Finally, we obtain the user's competing-aware interest $competing\_interest_t$ towards the target video $v_{target}$. The result in Table \ref{tab:rank_behav} shows that both the AUC and Logloss are improved by $0.005$ but the Recall and NDCG have no change. The result is not perfect as coarse ranking and it is worth exploring the modeling method continuously. 

\end{document}

%% file: math_commands.tex
\usepackage{amsmath,amsfonts,bm}

\def\eqref#1{equation~\ref{#1}}
\def\1{\bm{1}}

\DeclareMathAlphabet{\mathsfit}{\encodingdefault}{\sfdefault}{m}{sl}
\SetMathAlphabet{\mathsfit}{bold}{\encodingdefault}{\sfdefault}{bx}{n}

%% file: iclr2025_conference.bbl
\begin{thebibliography}{59}
\providecommand{\natexlab}[1]{#1}
\providecommand{\url}[1]{\texttt{#1}}
\expandafter\ifx\csname urlstyle\endcsname\relax
  \providecommand{\doi}[1]{doi: #1}\else
  \providecommand{\doi}{doi: \begingroup \urlstyle{rm}\Url}\fi

\bibitem[Asghar(2016)]{asghar2016yelp}
Nabiha Asghar.
\newblock Yelp dataset challenge: Review rating prediction.
\newblock \emph{arXiv preprint arXiv:1605.05362}, 2016.

\bibitem[Bai et~al.(2023)Bai, Jagerman, Qin, Yan, Kar, Lin, Wang, Bendersky, and Najork]{bai2023regression}
Aijun Bai, Rolf Jagerman, Zhen Qin, Le~Yan, Pratyush Kar, Bing-Rong Lin, Xuanhui Wang, Michael Bendersky, and Marc Najork.
\newblock Regression compatible listwise objectives for calibrated ranking with binary relevance.
\newblock In \emph{Proceedings of the 32nd ACM International Conference on Information and Knowledge Management}, pp.\  4502--4508, 2023.

\bibitem[Bello et~al.(2018)Bello, Kulkarni, Jain, Boutilier, Chi, Eban, Luo, Mackey, and Meshi]{bello2018seq2slate}
Irwan Bello, Sayali Kulkarni, Sagar Jain, Craig Boutilier, Ed~Chi, Elad Eban, Xiyang Luo, Alan Mackey, and Ofer Meshi.
\newblock Seq2slate: Re-ranking and slate optimization with rnns.
\newblock \emph{arXiv preprint arXiv:1810.02019}, 2018.

\bibitem[Bian et~al.(2022)Bian, Wu, Ren, Pi, Zhang, Xiao, Sheng, Zhu, Chan, Mou, et~al.]{bian2022can}
Weijie Bian, Kailun Wu, Lejian Ren, Qi~Pi, Yujing Zhang, Can Xiao, Xiang-Rong Sheng, Yong-Nan Zhu, Zhangming Chan, Na~Mou, et~al.
\newblock Can: feature co-action network for click-through rate prediction.
\newblock In \emph{Proceedings of the fifteenth ACM international conference on web search and data mining}, pp.\  57--65, 2022.

\bibitem[Chang et~al.(2023)Chang, Zhang, Fu, Zang, Guan, Lu, Hui, Leng, Niu, Song, et~al.]{chang2023twin}
Jianxin Chang, Chenbin Zhang, Zhiyi Fu, Xiaoxue Zang, Lin Guan, Jing Lu, Yiqun Hui, Dewei Leng, Yanan Niu, Yang Song, et~al.
\newblock Twin: Two-stage interest network for lifelong user behavior modeling in ctr prediction at kuaishou.
\newblock In \emph{Proceedings of the 29th ACM SIGKDD Conference on Knowledge Discovery and Data Mining}, pp.\  3785--3794, 2023.

\bibitem[Chen et~al.(2023)Chen, Dong, Wang, Feng, Wang, and He]{chen2023bias}
Jiawei Chen, Hande Dong, Xiang Wang, Fuli Feng, Meng Wang, and Xiangnan He.
\newblock Bias and debias in recommender system: A survey and future directions.
\newblock \emph{ACM Transactions on Information Systems}, 41\penalty0 (3):\penalty0 1--39, 2023.

\bibitem[Cheng et~al.(2016)Cheng, Koc, Harmsen, Shaked, Chandra, Aradhye, Anderson, Corrado, Chai, Ispir, et~al.]{cheng2016wide}
Heng-Tze Cheng, Levent Koc, Jeremiah Harmsen, Tal Shaked, Tushar Chandra, Hrishi Aradhye, Glen Anderson, Greg Corrado, Wei Chai, Mustafa Ispir, et~al.
\newblock Wide \& deep learning for recommender systems.
\newblock In \emph{Proceedings of the 1st workshop on deep learning for recommender systems}, pp.\  7--10, 2016.

\bibitem[Covington et~al.(2016)Covington, Adams, and Sargin]{covington2016deep}
Paul Covington, Jay Adams, and Emre Sargin.
\newblock Deep neural networks for youtube recommendations.
\newblock In \emph{Proceedings of the 10th ACM conference on recommender systems}, pp.\  191--198, 2016.

\bibitem[Deng et~al.(2009)Deng, Dong, Socher, Li, Li, and Fei-Fei]{deng2009imagenet}
Jia Deng, Wei Dong, Richard Socher, Li-Jia Li, Kai Li, and Li~Fei-Fei.
\newblock Imagenet: A large-scale hierarchical image database.
\newblock In \emph{2009 IEEE conference on computer vision and pattern recognition}, pp.\  248--255. Ieee, 2009.

\bibitem[Ding et~al.(2019)Ding, Quan, He, Li, and Jin]{ding2019reinforced}
Jingtao Ding, Yuhan Quan, Xiangnan He, Yong Li, and Depeng Jin.
\newblock Reinforced negative sampling for recommendation with exposure data.
\newblock In \emph{IJCAI}, pp.\  2230--2236. Macao, 2019.

\bibitem[Fan et~al.(2022)Fan, Ou, Gu, Fu, Li, Bao, Dai, Zeng, Zhuang, and Liu]{fan2022modeling}
Zhifang Fan, Dan Ou, Yulong Gu, Bairan Fu, Xiang Li, Wentian Bao, Xin-Yu Dai, Xiaoyi Zeng, Tao Zhuang, and Qingwen Liu.
\newblock Modeling users' contextualized page-wise feedback for click-through rate prediction in e-commerce search.
\newblock In \emph{Proceedings of the Fifteenth ACM International Conference on Web Search and Data Mining}, pp.\  262--270, 2022.

\bibitem[Gao et~al.(2022{\natexlab{a}})Gao, Li, Lei, Chen, Li, Jiang, He, Mao, and Chua]{gao2022kuairec}
Chongming Gao, Shijun Li, Wenqiang Lei, Jiawei Chen, Biao Li, Peng Jiang, Xiangnan He, Jiaxin Mao, and Tat-Seng Chua.
\newblock Kuairec: A fully-observed dataset and insights for evaluating recommender systems.
\newblock In \emph{Proceedings of the 31st ACM International Conference on Information \& Knowledge Management}, pp.\  540--550, 2022{\natexlab{a}}.

\bibitem[Gao et~al.(2022{\natexlab{b}})Gao, Li, Zhang, Chen, Li, Lei, Jiang, and He]{gao2022kuairand}
Chongming Gao, Shijun Li, Yuan Zhang, Jiawei Chen, Biao Li, Wenqiang Lei, Peng Jiang, and Xiangnan He.
\newblock Kuairand: an unbiased sequential recommendation dataset with randomly exposed videos.
\newblock In \emph{Proceedings of the 31st ACM International Conference on Information \& Knowledge Management}, pp.\  3953--3957, 2022{\natexlab{b}}.

\bibitem[Guo et~al.(2017)Guo, Tang, Ye, Li, and He]{guo2017deepfm}
Huifeng Guo, Ruiming Tang, Yunming Ye, Zhenguo Li, and Xiuqiang He.
\newblock Deepfm: a factorization-machine based neural network for ctr prediction.
\newblock \emph{arXiv preprint arXiv:1703.04247}, 2017.

\bibitem[Harper \& Konstan(2015)Harper and Konstan]{harper2015movielens}
F~Maxwell Harper and Joseph~A Konstan.
\newblock The movielens datasets: History and context.
\newblock \emph{Acm transactions on interactive intelligent systems (tiis)}, 5\penalty0 (4):\penalty0 1--19, 2015.

\bibitem[He et~al.(2014)He, Pan, Jin, Xu, Liu, Xu, Shi, Atallah, Herbrich, Bowers, et~al.]{he2014practical}
Xinran He, Junfeng Pan, Ou~Jin, Tianbing Xu, Bo~Liu, Tao Xu, Yanxin Shi, Antoine Atallah, Ralf Herbrich, Stuart Bowers, et~al.
\newblock Practical lessons from predicting clicks on ads at facebook.
\newblock In \emph{Proceedings of the eighth international workshop on data mining for online advertising}, pp.\  1--9, 2014.

\bibitem[Hidasi et~al.(2015)Hidasi, Karatzoglou, Baltrunas, and Tikk]{hidasi2015session}
Bal{\'a}zs Hidasi, Alexandros Karatzoglou, Linas Baltrunas, and Domonkos Tikk.
\newblock Session-based recommendations with recurrent neural networks.
\newblock \emph{arXiv preprint arXiv:1511.06939}, 2015.

\bibitem[Hou et~al.(2023)Hou, Wang, Liu, Qu, Cheng, and Lei]{hou2023deep}
Xuyang Hou, Zhe Wang, Qi~Liu, Tan Qu, Jia Cheng, and Jun Lei.
\newblock Deep context interest network for click-through rate prediction.
\newblock In \emph{Proceedings of the 32nd ACM International Conference on Information and Knowledge Management}, pp.\  3948--3952, 2023.

\bibitem[Huang et~al.(2013)Huang, He, Gao, Deng, Acero, and Heck]{huang2013learning}
Po-Sen Huang, Xiaodong He, Jianfeng Gao, Li~Deng, Alex Acero, and Larry Heck.
\newblock Learning deep structured semantic models for web search using clickthrough data.
\newblock In \emph{Proceedings of the 22nd ACM international conference on Information \& Knowledge Management}, pp.\  2333--2338, 2013.

\bibitem[Huang et~al.(2019)Huang, Zhang, and Zhang]{huang2019fibinet}
Tongwen Huang, Zhiqi Zhang, and Junlin Zhang.
\newblock Fibinet: combining feature importance and bilinear feature interaction for click-through rate prediction.
\newblock In \emph{Proceedings of the 13th ACM conference on recommender systems}, pp.\  169--177, 2019.

\bibitem[Imambi et~al.(2021)Imambi, Prakash, and Kanagachidambaresan]{imambi2021pytorch}
Sagar Imambi, Kolla~Bhanu Prakash, and GR~Kanagachidambaresan.
\newblock Pytorch.
\newblock \emph{Programming with TensorFlow: Solution for Edge Computing Applications}, pp.\  87--104, 2021.

\bibitem[Kang \& McAuley(2018)Kang and McAuley]{kang2018self}
Wang-Cheng Kang and Julian McAuley.
\newblock Self-attentive sequential recommendation.
\newblock In \emph{2018 IEEE international conference on data mining (ICDM)}, pp.\  197--206. IEEE, 2018.

\bibitem[Kingma \& Ba(2014)Kingma and Ba]{kingma2014adam}
Diederik~P Kingma and Jimmy Ba.
\newblock Adam: A method for stochastic optimization.
\newblock \emph{arXiv preprint arXiv:1412.6980}, 2014.

\bibitem[Li et~al.(2023)Li, Chen, Dong, Zhang, Wang, Wang, and Wang]{li2023decision}
Xiang Li, Shuwei Chen, Jian Dong, Jin Zhang, Yongkang Wang, Xingxing Wang, and Dong Wang.
\newblock Decision-making context interaction network for click-through rate prediction.
\newblock In \emph{Proceedings of the AAAI Conference on Artificial Intelligence}, volume~37, pp.\  5195--5202, 2023.

\bibitem[Lian et~al.(2020)Lian, Liu, and Chen]{lian2020personalized}
Defu Lian, Qi~Liu, and Enhong Chen.
\newblock Personalized ranking with importance sampling.
\newblock In \emph{Proceedings of The Web Conference 2020}, pp.\  1093--1103, 2020.

\bibitem[Lin et~al.(2023)Lin, Chen, Song, Liu, Li, and Jiang]{lin2023tree}
Xiao Lin, Xiaokai Chen, Linfeng Song, Jingwei Liu, Biao Li, and Peng Jiang.
\newblock Tree based progressive regression model for watch-time prediction in short-video recommendation.
\newblock In \emph{Proceedings of the 29th ACM SIGKDD Conference on Knowledge Discovery and Data Mining}, pp.\  4497--4506, 2023.

\bibitem[Lin et~al.(2024)Lin, Pan, Zhang, Wang, Xiao, Huang, Xiao, and Jiang]{lin2024understanding}
Zhutian Lin, Junwei Pan, Shangyu Zhang, Ximei Wang, Xi~Xiao, Shudong Huang, Lei Xiao, and Jie Jiang.
\newblock Understanding the ranking loss for recommendation with sparse user feedback.
\newblock \emph{arXiv preprint arXiv:2403.14144}, 2024.

\bibitem[Liu et~al.(2023)Liu, Zhou, Jiang, Ge, and Lian]{liu2023deep}
Qi~Liu, Zhilong Zhou, Gangwei Jiang, Tiezheng Ge, and Defu Lian.
\newblock Deep task-specific bottom representation network for multi-task recommendation.
\newblock In \emph{Proceedings of the 32nd ACM International Conference on Information and Knowledge Management}, pp.\  1637--1646, 2023.

\bibitem[Liu et~al.(2024)Liu, Hou, Lian, Wang, Jin, Cheng, and Lei]{liu2024at4ctr}
Qi~Liu, Xuyang Hou, Defu Lian, Zhe Wang, Haoran Jin, Jia Cheng, and Jun Lei.
\newblock At4ctr: Auxiliary match tasks for enhancing click-through rate prediction.
\newblock In \emph{Proceedings of the AAAI Conference on Artificial Intelligence}, volume~38, pp.\  8787--8795, 2024.

\bibitem[Lou et~al.(2022)Lou, Wen, Lv, Zhang, Yuan, and Li]{lou2022re}
Jiazhen Lou, Hong Wen, Fuyu Lv, Jing Zhang, Tengfei Yuan, and Zhao Li.
\newblock Re-weighting negative samples for model-agnostic matching.
\newblock In \emph{Proceedings of the 45th International ACM SIGIR Conference on Research and Development in Information Retrieval}, pp.\  1823--1827, 2022.

\bibitem[Ma et~al.(2018{\natexlab{a}})Ma, Zhao, Yi, Chen, Hong, and Chi]{ma2018modeling}
Jiaqi Ma, Zhe Zhao, Xinyang Yi, Jilin Chen, Lichan Hong, and Ed~H Chi.
\newblock Modeling task relationships in multi-task learning with multi-gate mixture-of-experts.
\newblock In \emph{Proceedings of the 24th ACM SIGKDD international conference on knowledge discovery \& data mining}, pp.\  1930--1939, 2018{\natexlab{a}}.

\bibitem[Ma et~al.(2018{\natexlab{b}})Ma, Zhao, Huang, Wang, Hu, Zhu, and Gai]{ma2018entire}
Xiao Ma, Liqin Zhao, Guan Huang, Zhi Wang, Zelin Hu, Xiaoqiang Zhu, and Kun Gai.
\newblock Entire space multi-task model: An effective approach for estimating post-click conversion rate.
\newblock In \emph{The 41st International ACM SIGIR Conference on Research \& Development in Information Retrieval}, pp.\  1137--1140, 2018{\natexlab{b}}.

\bibitem[Ni et~al.(2019)Ni, Li, and McAuley]{ni2019justifying}
Jianmo Ni, Jiacheng Li, and Julian McAuley.
\newblock Justifying recommendations using distantly-labeled reviews and fine-grained aspects.
\newblock In \emph{Proceedings of the 2019 conference on empirical methods in natural language processing and the 9th international joint conference on natural language processing (EMNLP-IJCNLP)}, pp.\  188--197, 2019.

\bibitem[Pei et~al.(2019)Pei, Zhang, Zhang, Sun, Lin, Sun, Wu, Jiang, Ge, Ou, et~al.]{pei2019personalized}
Changhua Pei, Yi~Zhang, Yongfeng Zhang, Fei Sun, Xiao Lin, Hanxiao Sun, Jian Wu, Peng Jiang, Junfeng Ge, Wenwu Ou, et~al.
\newblock Personalized re-ranking for recommendation.
\newblock In \emph{Proceedings of the 13th ACM conference on recommender systems}, pp.\  3--11, 2019.

\bibitem[Qin et~al.(2022)Qin, Zhu, Chen, Liu, Liu, Tang, Zhang, Yu, and Zhang]{qin2022rankflow}
Jiarui Qin, Jiachen Zhu, Bo~Chen, Zhirong Liu, Weiwen Liu, Ruiming Tang, Rui Zhang, Yong Yu, and Weinan Zhang.
\newblock Rankflow: Joint optimization of multi-stage cascade ranking systems as flows.
\newblock In \emph{Proceedings of the 45th International ACM SIGIR Conference on Research and Development in Information Retrieval}, pp.\  814--824, 2022.

\bibitem[Rendle \& Freudenthaler(2014)Rendle and Freudenthaler]{rendle2014improving}
Steffen Rendle and Christoph Freudenthaler.
\newblock Improving pairwise learning for item recommendation from implicit feedback.
\newblock In \emph{Proceedings of the 7th ACM international conference on Web search and data mining}, pp.\  273--282, 2014.

\bibitem[Rendle et~al.(2012)Rendle, Freudenthaler, Gantner, and Schmidt-Thieme]{rendle2012bpr}
Steffen Rendle, Christoph Freudenthaler, Zeno Gantner, and Lars Schmidt-Thieme.
\newblock Bpr: Bayesian personalized ranking from implicit feedback.
\newblock \emph{arXiv preprint arXiv:1205.2618}, 2012.

\bibitem[Sheng et~al.(2023)Sheng, Gao, Cheng, Yang, Han, Deng, Jiang, Xu, and Zheng]{sheng2023joint}
Xiang-Rong Sheng, Jingyue Gao, Yueyao Cheng, Siran Yang, Shuguang Han, Hongbo Deng, Yuning Jiang, Jian Xu, and Bo~Zheng.
\newblock Joint optimization of ranking and calibration with contextualized hybrid model.
\newblock In \emph{Proceedings of the 29th ACM SIGKDD Conference on Knowledge Discovery and Data Mining}, pp.\  4813--4822, 2023.

\bibitem[Shi et~al.(2019)Shi, Yu, Da, Chen, and Zeng]{shi2019virtual}
Jing-Cheng Shi, Yang Yu, Qing Da, Shi-Yong Chen, and An-Xiang Zeng.
\newblock Virtual-taobao: Virtualizing real-world online retail environment for reinforcement learning.
\newblock In \emph{Proceedings of the AAAI Conference on Artificial Intelligence}, volume~33, pp.\  4902--4909, 2019.

\bibitem[Song et~al.(2022)Song, Huang, Wang, Huang, Yu, Chen, Yao, Fan, Peng, Lin, et~al.]{song2022rethinking}
Jinbo Song, Ruoran Huang, Xinyang Wang, Wei Huang, Qian Yu, Mingming Chen, Yafei Yao, Chaosheng Fan, Changping Peng, Zhangang Lin, et~al.
\newblock Rethinking large-scale pre-ranking system: Entire-chain cross-domain models.
\newblock In \emph{Proceedings of the 31st ACM International Conference on Information \& Knowledge Management}, pp.\  4495--4499, 2022.

\bibitem[Sun et~al.(2023)Sun, Si, Zang, Leng, Niu, Song, Zhang, and Xu]{sun2023kuaisar}
Zhongxiang Sun, Zihua Si, Xiaoxue Zang, Dewei Leng, Yanan Niu, Yang Song, Xiao Zhang, and Jun Xu.
\newblock Kuaisar: A unified search and recommendation dataset.
\newblock In \emph{Proceedings of the 32nd ACM International Conference on Information and Knowledge Management}, pp.\  5407--5411, 2023.

\bibitem[Tang et~al.(2020)Tang, Liu, Zhao, and Gong]{tang2020progressive}
Hongyan Tang, Junning Liu, Ming Zhao, and Xudong Gong.
\newblock Progressive layered extraction (ple): A novel multi-task learning (mtl) model for personalized recommendations.
\newblock In \emph{Fourteenth ACM Conference on Recommender Systems}, pp.\  269--278, 2020.

\bibitem[Wang et~al.(2018)Wang, Singh, Michael, Hill, Levy, and Bowman]{wang2018glue}
Alex Wang, Amanpreet Singh, Julian Michael, Felix Hill, Omer Levy, and Samuel~R Bowman.
\newblock Glue: A multi-task benchmark and analysis platform for natural language understanding.
\newblock \emph{arXiv preprint arXiv:1804.07461}, 2018.

\bibitem[Wang et~al.(2022)Wang, Wang, Li, Gu, Lu, Zhang, and Gu]{wang2022enhancing}
Fangye Wang, Yingxu Wang, Dongsheng Li, Hansu Gu, Tun Lu, Peng Zhang, and Ning Gu.
\newblock Enhancing ctr prediction with context-aware feature representation learning.
\newblock In \emph{Proceedings of the 45th International ACM SIGIR Conference on Research and Development in Information Retrieval}, pp.\  343--352, 2022.

\bibitem[Wang et~al.(2020)Wang, Zhao, Jiang, Zhou, Zhu, and Gai]{wang2020cold}
Zhe Wang, Liqin Zhao, Biye Jiang, Guorui Zhou, Xiaoqiang Zhu, and Kun Gai.
\newblock Cold: Towards the next generation of pre-ranking system.
\newblock \emph{arXiv preprint arXiv:2007.16122}, 2020.

\bibitem[Wei et~al.(2024)Wei, Zhou, Wu, and Liu]{wei2024enhancing}
Jianping Wei, Yujie Zhou, Zhengwei Wu, and Ziqi Liu.
\newblock Enhancing pre-ranking performance: Tackling intermediary challenges in multi-stage cascading recommendation systems.
\newblock In \emph{Proceedings of the 30th ACM SIGKDD Conference on Knowledge Discovery and Data Mining}, pp.\  5950--5958, 2024.

\bibitem[Yan et~al.(2022)Yan, Qin, Wang, Bendersky, and Najork]{yan2022scale}
Le~Yan, Zhen Qin, Xuanhui Wang, Michael Bendersky, and Marc Najork.
\newblock Scale calibration of deep ranking models.
\newblock In \emph{Proceedings of the 28th ACM SIGKDD Conference on Knowledge Discovery and Data Mining}, pp.\  4300--4309, 2022.

\bibitem[Yuan et~al.(2022)Yuan, Yuan, Li, Kong, Li, Chen, Yang, Yu, Hu, Li, et~al.]{yuan2022tenrec}
Guanghu Yuan, Fajie Yuan, Yudong Li, Beibei Kong, Shujie Li, Lei Chen, Min Yang, Chenyun Yu, Bo~Hu, Zang Li, et~al.
\newblock Tenrec: A large-scale multipurpose benchmark dataset for recommender systems.
\newblock \emph{Advances in Neural Information Processing Systems}, 35:\penalty0 11480--11493, 2022.

\bibitem[Zhang et~al.(2013)Zhang, Chen, Wang, and Yu]{zhang2013optimizing}
Weinan Zhang, Tianqi Chen, Jun Wang, and Yong Yu.
\newblock Optimizing top-n collaborative filtering via dynamic negative item sampling.
\newblock In \emph{Proceedings of the 36th international ACM SIGIR conference on Research and development in information retrieval}, pp.\  785--788, 2013.

\bibitem[Zhang et~al.(2023{\natexlab{a}})Zhang, Dong, Ding, Li, Jiang, and Gai]{zhang2023divide}
Yuan Zhang, Xue Dong, Weijie Ding, Biao Li, Peng Jiang, and Kun Gai.
\newblock Divide and conquer: Towards better embedding-based retrieval for recommender systems from a multi-task perspective.
\newblock In \emph{Companion Proceedings of the ACM Web Conference 2023}, pp.\  366--370, 2023{\natexlab{a}}.

\bibitem[Zhang et~al.(2022)Zhang, Sheng, Zhang, Jiang, Han, Deng, and Zheng]{zhang2022towards}
Zhao-Yu Zhang, Xiang-Rong Sheng, Yujing Zhang, Biye Jiang, Shuguang Han, Hongbo Deng, and Bo~Zheng.
\newblock Towards understanding the overfitting phenomenon of deep click-through rate models.
\newblock In \emph{Proceedings of the 31st ACM international conference on information \& knowledge management}, pp.\  2671--2680, 2022.

\bibitem[Zhang et~al.(2023{\natexlab{b}})Zhang, Huang, Ou, Li, Li, Liu, and Zeng]{zhang2023rethinking}
Zhixuan Zhang, Yuheng Huang, Dan Ou, Sen Li, Longbin Li, Qingwen Liu, and Xiaoyi Zeng.
\newblock Rethinking the role of pre-ranking in large-scale e-commerce searching system.
\newblock \emph{arXiv preprint arXiv:2305.13647}, 2023{\natexlab{b}}.

\bibitem[Zhao et~al.(2023)Zhao, Liu, Cai, Zhao, Liu, Zheng, Jiang, and Gai]{zhao2023kuaisim}
Kesen Zhao, Shuchang Liu, Qingpeng Cai, Xiangyu Zhao, Ziru Liu, Dong Zheng, Peng Jiang, and Kun Gai.
\newblock Kuaisim: A comprehensive simulator for recommender systems.
\newblock \emph{Advances in Neural Information Processing Systems}, 36:\penalty0 44880--44897, 2023.

\bibitem[Zhao et~al.(2019)Zhao, Hong, Wei, Chen, Nath, Andrews, Kumthekar, Sathiamoorthy, Yi, and Chi]{zhao2019recommending}
Zhe Zhao, Lichan Hong, Li~Wei, Jilin Chen, Aniruddh Nath, Shawn Andrews, Aditee Kumthekar, Maheswaran Sathiamoorthy, Xinyang Yi, and Ed~Chi.
\newblock Recommending what video to watch next: a multitask ranking system.
\newblock In \emph{Proceedings of the 13th ACM Conference on Recommender Systems}, pp.\  43--51, 2019.

\bibitem[Zheng et~al.(2024)Zheng, Zhao, Huang, Zhang, Mou, Niu, Song, Wang, and Gai]{zheng2024full}
Kai Zheng, Haijun Zhao, Rui Huang, Beichuan Zhang, Na~Mou, Yanan Niu, Yang Song, Hongning Wang, and Kun Gai.
\newblock Full stage learning to rank: A unified framework for multi-stage systems.
\newblock In \emph{Proceedings of the ACM on Web Conference 2024}, pp.\  3621--3631, 2024.

\bibitem[Zheng et~al.(2022)Zheng, Wang, Li, Chen, Liu, Lu, Zhao, Peng, Lin, and Shao]{zheng2022implicit}
Kaifu Zheng, Lu~Wang, Yu~Li, Xusong Chen, Hu~Liu, Jing Lu, Xiwei Zhao, Changping Peng, Zhangang Lin, and Jingping Shao.
\newblock Implicit user awareness modeling via candidate items for ctr prediction in search ads.
\newblock In \emph{Proceedings of the ACM Web Conference 2022}, pp.\  246--255, 2022.

\bibitem[Zhou et~al.(2018)Zhou, Zhu, Song, Fan, Zhu, Ma, Yan, Jin, Li, and Gai]{zhou2018deep}
Guorui Zhou, Xiaoqiang Zhu, Chenru Song, Ying Fan, Han Zhu, Xiao Ma, Yanghui Yan, Junqi Jin, Han Li, and Kun Gai.
\newblock Deep interest network for click-through rate prediction.
\newblock In \emph{Proceedings of the 24th ACM SIGKDD international conference on knowledge discovery \& data mining}, pp.\  1059--1068, 2018.

\bibitem[Zhou et~al.(2019)Zhou, Mou, Fan, Pi, Bian, Zhou, Zhu, and Gai]{zhou2019deep}
Guorui Zhou, Na~Mou, Ying Fan, Qi~Pi, Weijie Bian, Chang Zhou, Xiaoqiang Zhu, and Kun Gai.
\newblock Deep interest evolution network for click-through rate prediction.
\newblock In \emph{Proceedings of the AAAI conference on artificial intelligence}, volume~33, pp.\  5941--5948, 2019.

\bibitem[Zhu et~al.(2018)Zhu, Li, Zhang, Li, He, Li, and Gai]{zhu2018learning}
Han Zhu, Xiang Li, Pengye Zhang, Guozheng Li, Jie He, Han Li, and Kun Gai.
\newblock Learning tree-based deep model for recommender systems.
\newblock In \emph{Proceedings of the 24th ACM SIGKDD International Conference on Knowledge Discovery \& Data Mining}, pp.\  1079--1088, 2018.

\end{thebibliography}
